\documentclass{article}
\usepackage[preprint]{neurips_2023}

\usepackage{booktabs}

\newcommand{\mathbbm}[1]{\text{\usefont{U}{bbm}{m}{n}#1}} 
\usepackage{subcaption}
\usepackage{graphicx}

\usepackage{soul}
\usepackage{transparent}
\usepackage[normalem]{ulem}
\useunder{\uline}{\ul}{}

\usepackage{enumitem}
\setlist[itemize]{leftmargin=5mm}

\usepackage{pgfplots}
\pgfplotsset{compat=1.17}
\usepackage{colortbl}
\usepackage{xcolor}
\usepackage{pgfplotstable}
\usepackage{caption}
\usepackage{wrapfig}
\usepackage{amsmath, amsthm, amssymb, amsfonts}
\usepackage{url}

\AtBeginDocument{%
  }

\begin{document}

\title{RELand: Risk Estimation of Landmines via Interpretable Invariant Risk Minimization}

\author{%
  Mateo Dulce Rubio$^{*,1}$, Siqi Zeng$^{*,1}$, Qi Wang$^{1}$, Didier Alvarado$^{2}$, Francisco Moreno$^{3}$,\\
  \textbf{Hoda Heidari$^{1}$, Fei Fang$^{1}$} \\~\\
    $^1$ Carnegie Mellon University \\
    $^2$ United Nations Mine Action Service \\
    $^3$ Colombian Campaign to Ban Landmines \\~\\
Corresponding author: \texttt{mdulceru@andrew.cmu.edu} \\
$^*$Both authors contributed equally to this research.
}

\maketitle

\begin{abstract}
  Landmines remain a threat to war-affected communities for years after conflicts have ended, partly due to the laborious nature of demining tasks. Humanitarian demining operations begin by collecting relevant information from the sites to be cleared, which is then analyzed by human experts to determine the potential risk of remaining landmines. In this paper, we propose \textbf{\textsc{RELand}} system to support these tasks, which consists of three major components. We (1) provide general feature engineering and label assigning guidelines to enhance datasets for landmine risk modeling, which are widely applicable to global demining routines, (2) formulate landmine presence as a classification problem and design a novel interpretable model based on sparse feature masking and invariant risk minimization, and run extensive evaluation under proper protocols that resemble real-world demining operations to show a significant improvement over the state-of-the-art, and (3) build an interactive web interface to suggest priority areas for demining organizations. We are currently collaborating with a humanitarian demining NGO in Colombia that is using our system as part of their field operations in two areas recently prioritized for demining.
\end{abstract}

\section{Introduction}

\begin{wrapfigure}{r}{0.2\linewidth}
\begin{center}
    \includegraphics[width=0.17\textwidth]{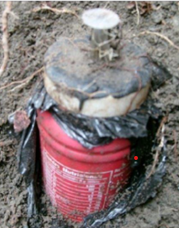}
\end{center}
\caption{A landmine in Colombia \citep{Exampleofalandmine}.}
\label{fig:mine}
\end{wrapfigure}

Anti-personnel landmines (Fig. \ref{fig:mine}) are explosive devices hidden in the ground designed to explode by proximity or contact and with the capacity to kill, disable or cause harm to humans \citep{UNprotocol, gichd_2009}. The mere threat of landmines has negative consequences on war-affected communities, including the loss of forest areas \citep{forest}, reduction of productive land \citep{KONIUSZEWSKI2016119, Trevelyan2002FarmingM}, exacerbation of social vulnerability \citep{social-vulnerability}, delay of infrastructure development \citep{doi:10.1177/15423166211068324}, and damage of natural, physical, and social capital \citep{nonconventional}, among others. Due to such consequences, in 1997 most countries signed the Ottawa Treaty committing themselves to stop the manufacture, commercialization, and use of antipersonnel landmines. Likewise, the countries that had historically used these weapons during armed conflicts undertook to clear their territories. Despite the efforts, these explosive devices are still used in different conflicts around the world and remain a threat to humanity, with millions of them still not found across 60 countries \citep{local-effect}.

Humanitarian demining operations seek to clear war-affected regions of possible antipersonnel mines so that communities can safely reland the areas. According to international standards, these operations are divided into three main phases \citep{gichd_2009, nonTechnical, Exampleofalandmine}. First, there is the \textit{non-technical survey} in which humanitarian demining organizations use official data and collect information through surveys with the residents in a territory about the possible presence of antipersonnel mines to define initial hazardous areas. Second, there is the \textit{technical survey}, which involves technical interventions to collect further information (e.g., manual, mechanical or animal detection mechanisms) to confirm or cancel the previously identified hazardous areas \citep{gichd_2009}. Finally, in the confirmed hazard areas, the \textit{clearance} phase takes place in which the actual clearance of the mines from the ground is carried out. International standards for mine clearance mandate a comprehensive inspection of the suspected areas before declaring them safe. As a consequence, demining operations are laborious and costly. 

Our work is primarily proposed to support non-technical surveys in humanitarian demining. To the best of our understanding, this phase is currently carried out by experts in the territories who evaluate the possible presence of antipersonnel mines based on available information and that provided by the communities that inhabit such areas. 
The past presence of armed groups, proximity to areas of illegal economies, or historic antipersonnel mine events nearby may suggest the use of landmines in a given geographical region \citep{cnmh}. 
Since landmines are not used randomly but under war logic, Machine Learning (ML) based systems can potentially help with these surveys by analyzing the historical landmine events and their correlation to relevant features.

However, estimating the risk of antipersonnel landmine presence poses three main challenges. Firstly, false negatives, which indicate the absence of mines when they are actually present, pose a greater risk to human lives compared to false positives, which incorrectly indicate the presence of mines and lead to resource misallocation. Secondly, the problem itself is highly imbalanced, resulting in conventional models generating a high rate of false alarms and, consequently, making their use inefficient. Finally, humanitarian demining operations are in general carried out in areas that have not been previously investigated due to security reasons, and which are subject to arbitrary distribution shifts with respect to the available training data from previously cleared areas. Previous attempts to address these challenges using machine learning (ML) models in demining operations have key limitations. Some studies used a limited set of features \citep{corredorUNAL, unpublishedCAG} or relied on hard-to-collect data \citep{whaite_uniandes,curriculum,tiramisu,radar}, which hinders the usefulness of the models. Some have used misspecified labels  \citep{CAG_paper, density_mapping_alegria} or not well-justified evaluation protocols \citep{baseline, aspatial}, undermining the credibility of their results. Overall, formulating the presence of antipersonnel mines as a standard classification problem proves inadequate. 

\begin{figure}
    \centering
    \includegraphics[width=\textwidth]{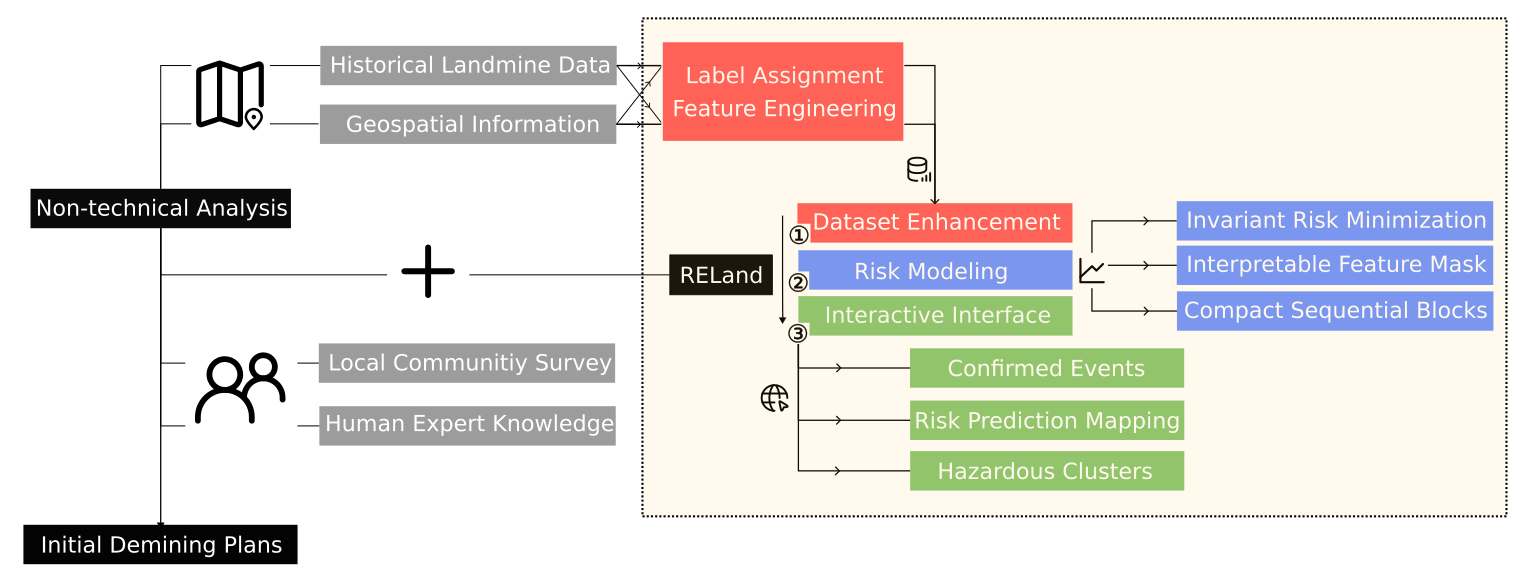}
    \caption{Integration of \textbf{\textsc{RELand}} system into the humanitarian demining pipeline. Current non-technical surveys (grey) are based on the visual inspection of data in geospatial information systems and human expert analyses including local community surveys and domain knowledge. \textbf{\textsc{RELand}} (yellow dashed box) serves as an additional toolbox that contains three major components: dataset enhancement based on existing public geospatial datasets (red), risk modeling with machine learning methods (blue), and interactive web interface (green).}
    \label{fig:pipeline}
\end{figure}

In this paper, we present the \textbf{\textsc{RELand}} system (Risk Estimation of Landmines), a sensible comprehensive pipeline to estimate the risk of landmine presence  designed to support non-technical surveys in humanitarian demining operations to define initial hazardous areas. We address the challenges of antipersonnel landmine risk estimation by enhancing existing datasets to include rich relevant features previously ignored in the literature, constructing a novel, robust, and interpretable ML model that outperforms both standard and new baselines, and developing an interactive web application together with key stakeholders to validate our approach. These three major components of \textbf{\textsc{RELand}} are detailed below and illustrated in Fig. \ref{fig:pipeline}. Notably, our approach is the first public pipeline of its kind that can be easily adapted for use in demining workflows globally.

\begin{enumerate}
    \item \textbf{Dataset Enhancement:} This component integrates different sources of information to construct a dataset for landmine presence with rich relevant features based on geographic information, socio-demographic variables, remnants of war indicators, and historical antipersonnel landmine events. We introduce several new features which prove useful to identify hazard areas and to rule out false alarms. We also argue how labels should be assigned to predict the results of humanitarian demining operations, rectifying the definition of labels in the literature. We describe this component in Section \ref{module_data}. 

    \item \textbf{Risk Modeling:} We designed a novel interpretable deep learning tabular model with Invariant Risk Minimization (IRM) \citep{InvariantRiskMinimization}, the core part of \textbf{\textsc{RELand}}, to generate risk prediction estimates. With sparse feature masking, our method offers convincing global feature importance verified by visualization of prediction maps and discussions with domain experts. In addition, the use of IRM makes our model robust to distribution shifts and perform equally well on ``easy'' areas with historical information about past landmine events, as well as on ``hard'' areas with limited or no historical recorded events. The proposed methodology is interesting in its own right and can be directly extended to other settings. We discuss the proposed methodology in Section \ref{module_model}. 

    \item \textbf{Interactive Interface:} We built a web application that presents the model output to support demining tasks and the efficient allocation of demining resources. Our interactive web interface comprises three layers of information: confirmed historical events, risk prediction maps, and identified hazardous clusters. The interface was also used to validate \textsc{\textbf{RELand}} approach by potential end users of the system. We detail the functionalities of the interactive interface in Section \ref{module_app}.
\end{enumerate}

In Section \ref{experiments}, we demonstrate the capability of the proposed system to support non-technical surveys in estimating the risk of antipersonnel landmine presence and identifying priority areas for demining. We run extensive experiments with real data from Colombia, the country with the second-largest number of victims of anti-personnel mines \citep{local-effect}, under simulated deployment scenarios. Using spatial cross-validation, we show that our model outperforms standard and new baselines across different metrics relevant to demining operations. Moreover, we are currently conducting a pilot field study (Section \ref{field_test}) in two geographic regions in Colombia recently allocated for humanitarian demining in partnership with a non-governmental organization and an intergovernmental organization, with initial positive feedback. Section \ref{conclusions} concludes and discusses the limitations and next steps of our work.

In addition to the \textbf{\textsc{RELand}} system, we contribute to the literature by designing validation protocols for ML models for humanitarian demining that appropriately quantifies model performance. We also extensively developed new and well-motivated baseline models for resource allocation problems that directly minimize metrics from the ranking learning literature \citep{rudin2009p}. Finally, our interpretable model can be used as part of Explosive Ordnance Risk Education (EORE), an essential pillar of comprehensive mine action \citep{AICMA}, to inform communities about possible suggestive evidence of the potential presence of antipersonnel mines in their territories. The resulting dataset and developed code are released for reproducibility and extension of our work.

\section{Related Work}
\paragraph{\textbf{ML Methods for Landmine Risk Modeling}} \ \ 
Previous studies have attempted to model landmine presence using supervised learning approaches, but face key limitations. For instance, \citet{baseline} proposed a landmine prediction model based on logistic regression (LR) and support vector machine (SVM), and \citet{aspatial} employed LR with and without spatial weights to generate landmine risk maps in localized regions in Bosnia and Herzegovina, and Colombia. However, these approaches \citep{baseline, aspatial} have overlooked key relevant factors associated with landmine use such as the presence of illegal economies, combats between the armies or rich historical landmine events information \citep{buschschluter_2010, cnmh}. Alternatively, other works have used data that requires expensive equipment, e.g., airborne images \citep{curriculum,tiramisu} or radar underground images \citep{radar}, which limits the feasibility of large-scale tests or deployment.

Moreover, a crucial problem that has been neglected in previous literature is the presence of ``noisy labels'' in the available data \citep{CAG_paper,density_mapping_alegria, baseline, aspatial}. Specifically, the absence of landmine events can either indicate the actual absence of landmines (i.e., a negative label) or simply that the landmines in that particular area have not yet been discovered. This lack of well-defined labels significantly impacts the validity of supervised learning models and raises concerns about the credibility of previously proposed systems. In addition, there is no standard evaluation protocol for landmine risk estimation models, and some previous works have only reported train performance under difficult-to-meet i.i.d. assumptions \citep{baseline, aspatial}. Ignoring the spatial correlation of the data in cross-validation approaches overestimates the capabilities of the proposed models. 

Finally, some studies have conducted unsupervised descriptive analyses of landmine presence using kernel density estimation \citep{density_mapping_alegria, Hazard, CAG_paper} applied directly to the historical events, or combining it with hierarchical clustering and remote sensing methods \citep{Combined}. Our focus is on addressing a predictive problem where unsupervised analyses cannot be easily applied and have limited inference power on unseen test regions. Furthermore, in our work, we argue for the correct way to assign labels, construct a rich dataset, and design a principled evaluation protocol for ML models aimed to estimate the risk of antipersonnel landmine presence.

\paragraph{\textbf{Deep Tabular Models and OOD Generalization Methods}} \ 
Deep learning models have achieved remarkable success in various machine learning applications for tabular datasets \citep{badirli2022gradient, klambauer2017self, TAHERKHANI2020351}. These models are generally built on overparametrized artificial neural networks that enable them to extract meaningful features from large and complex datasets. However, it is more challenging to interpret these black-box models with huge amounts of parameters compared to shallow models. This has led to the development of interpretable deep tabular methods in recent years. For example, TabNet \citep{tabnet} learns instance-wise feature importance with attention transformer modules, \citet{10.1145/3357384.3357925} studies feature interaction with multihead-attention layers, and \cite{KUMARI2021123285, SARKAR2022200097} both combine gradient boosting decision trees (GBDT) \citep{gbdt10.1214/aos/1013203451, xgboost} with deep neural networks and report feature importance based on information gain. In this paper, we avoid introducing empirically sub-optimal gradient boosting tree learners \citep{gbdt10.1214/aos/1013203451}, and therefore mainly study the extension of TabNet to generate global feature importance. 

Nevertheless, Out-of-Distribution (OOD) generalization remains a fundamental problem for Machine Learning models. This is particularly important for ML models that are deployed in the real world, where they may encounter test data whose distribution is arbitrarily different from the available training datasets. Despite some recent progress in the field, none of the popular domain generalization methods outperforms classic empirical risk minimization (ERM) consistently on image datasets \citep{ye2022oodbench}. Moreover, the performance comparison between ERM and OOD methods is inconsistent across different data modalities \citep{2022arXiv220109637J, WOODS, 10.1007/978-3-030-92659-5_39}, suggesting that the successful application of OOD inference methods depends on the type of distribution shift and domain-specific knowledge. In our problem, we show that the Invariant Risk Minimization 
approach proposed by \citet{InvariantRiskMinimization} consistently outperforms ERM-based baseline models.

\section{Dataset Enhancement}
\label{module_data}

In this section, we detail the data enhancement component of \textsc{\textbf{RELand}}. We focus on Colombia as a case study to illustrate the proposed methodology for assigning reliable labels that capture the results of humanitarian demining operations. We then make a systematic study of potentially relevant variables to estimate the risk of landmine presence. To our best knowledge, our constructed dataset is the first publicly available of its kind. 

\subsection{Case study: Colombia}

During the decades-long armed conflict in Colombia, antipersonnel mines were widely used as a means of achieving war purposes: illegal armed groups used landmines to protect areas of their interest, such as cocaine crops, laboratories, drug trafficking routes, and illegal mining sites, and as a combat strategy against war rivals \citep{buschschluter_2010, cnmh}. As a consequence, between 1990 and 2022, about 38,000 landmine events were recorded, resulting in more than 12,000 victims, about 20\% of them fatal \citep{departamento}. To confront this affliction, Colombia was one of the countries that signed the Ottawa Treaty committing to clear the country of antipersonnel mines by 2021 (now extended to 2025). Importantly, landmines are not uniformly distributed: 50\% of historical victims concentrate in nearly 2\% of the country's municipalities \citep{cnmh} and its use is strongly correlated with the presence of illegal economies such as coca crops and the presence of guerrillas, as documented in \cite{cnmh}. For a thorough summary of Colombian demining efforts see \cite{local-effect}.

We focus on the Antioquia region of Colombia for four main reasons: (i) Antioquia is the department with the highest historical number of landmine events and victims \citep{datasource1}, (ii) Data on historical landmine events, results of humanitarian demining operations and relevant variables for their prediction are freely available and of high quality; (iii) It has been used in previous works \citep{aspatial, baseline, CAG_paper}, which allows for a fair comparison between the proposed method and previous models in the literature; and (iv) Is currently conducting humanitarian demining operations in its territory and early conversations with some NGOs leading these operations evidenced the potential benefit of developing a system such as \textbf{\textsc{RELand}}. The patterns of spatial concentration of antipersonnel mines and their correlation with remnants of war indicators have been similarly documented in other contexts such as Bosnia-Herzegovina \citep{aspatial} and Cambodia, and we believe that the proposed data collection and modeling methodology can be directly extended to other countries where the information is available.

\begin{figure}
    \centering
    \includegraphics[width=.6\textwidth]{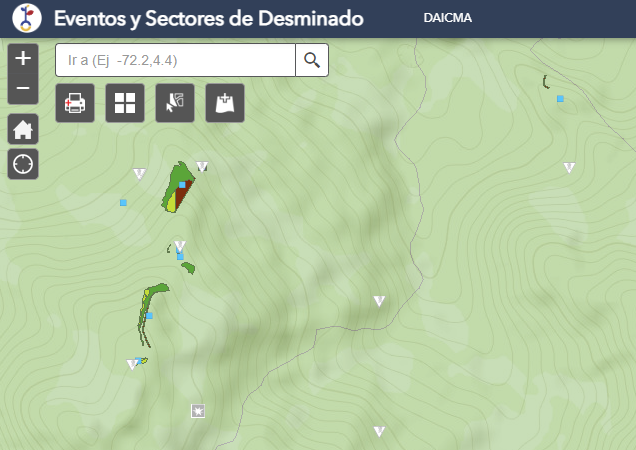}
    \caption{A screenshot of IMSMA-AICMA public web interface \citep{datasource1}. Historical events (grey points) and areas identified during humanitarian demining operations (red and green polygons).}
    \label{fig:IMSMA_DAICMA}
\end{figure}

As well as other countries, Colombia adopted the Information Management System for Mine Action (IMSMA) as part of its comprehensive anti-personnel mine action. IMSMA systematically records georeferenced historical landmine events since 1990 in the country reported by civilians and the military \citep{departamento}. The database is freely accessible and is the main input used by government agencies and humanitarian demining organizations to plan their operations. In addition, the Colombian National Mine Action Center (AICMA, for its Spanish initials) of the Colombian High Commissioner for Peace collects the results of demining operations carried out by non-governmental and military organizations \citep{datasource1}. The information is made public through their spatial viewer that presents hazardous, confirmed hazardous, canceled, and cleared areas as a result of humanitarian demining operations. Fig. \ref{fig:IMSMA_DAICMA} presents, for a small area in Antioquia, Colombia, the historical landmine events recorded in IMSMA (georeferenced points) and the areas identified by humanitarian demining organizations in their operations in the region (georeferenced polygons). To summarize, demining organizations use IMSMA data to plan their operations resulting in the identified areas reported in the AICMA viewer. 

We argue that reliable labels must reflect this temporal workflow: demining organizations have access to the historical landmine events and use this information to plan their operations, and conduct their non-technical surveys and subsequent interventions. Accordingly, we use historical events as one of the main variables to predict the areas found with landmines as a result of humanitarian demining operations. More details on the outcome definition and predictive variables are given below.

\subsection{Label Assignment}

Our main study region is composed by 15 municipalities in Antioquia that have been declared mine-free areas after comprehensive mine action operations.\footnote{Abejorral, Alejandría, Chigorodó, Cocorná, El Carmen de Viboral, Granada, La Unión, Nariño, Sabanalarga, San Carlos, San Francisco, San Luis, San Rafael, San Roque, Sonsón.} We divide each municipality into 500 by 500 meters homogeneous cells. A grid cell is assigned a positive label if it intersects at least one confirmed hazard area or a cleared area where landmines were found during humanitarian demining operations, using the information from AICMA \citep{datasource1}. Otherwise, a grid cell is assigned a negative label, given that these areas have been thoroughly inspected and declared free of mines. Our final dataset for the Antioquia region has only 1.6\% of positive grid cells, making the classification problem very difficult in nature.



\subsection{Feature Engineering}
Inspired by previous works we use three types of features relevant to estimate the risk of landmines presence \citep{Hazard, baseline, aspatial, corredorUNAL, unpublishedCAG}: (1) remnants of war indicators, (2) geographical features, and (3) socio-demographic characteristics. In addition to features used in prior literature, we carefully study many possible factors that may influence landmine locations and include them in our dataset for the first time among related work. This data collection process was discussed and complemented with civil and military demining organizations in Colombia. We use rich geographically disaggregated data to assign each cell with different attributes in each of the three categories and reliably estimate the presence of antipersonnel mines with the desired granularity. Our rich data sources also allow us to encode the spatial dynamics of landmines as both binary indicators and distances to points of interest.

Remnants of war indicators refer to variables that suggest the past presence of guerrillas and other armed groups involved in the Colombian armed conflict that potentially used antipersonnel mines. In this category, we include the distance to cocaine crops \citep{datasource7}, signs of deforestation \citep{doi:10.1126/science.1244693}, mining \citep{colombia-mining}, and power, telecommunication, and oil lines \citep{datasource9} (frequently attacked by guerrillas). The geographic variables capture characteristics of the terrain and suggest its potential use for criminal or agricultural activities and include altitude \citep{world-climate}, animal density \citep{threatened-species}, soil texture \citep{essd-14-4719-2022}, temperature \citep{world-climate}, rainfall \citep{world-climate}, distance to rivers \citep{datasource9}, and several categories of land use, weather, and topographic relief \citep{datasource5}. Finally, socio-demographic variables provide information on the communities that inhabit these sites and include the distance to airports, seaports, settlements, financial institutions, schools, buildings \citep{datasource9}, and roads \citep{roads_IGAC}, a relative wealth index \citep{datasource3}, population \citep{world-pop}, and a night lights index \citep{li2020harmonized} as a metric of economic development. In addition, we add municipality-level covariates to each grid cell including the total number of historical landmine events \citep{departamento}, number of victims during the armed conflict \citep{datasource6}, population, rurality index, distance to the capital of the department, per-capita GDP (2009 constant prices), average altitude and total area \citep{panel_CEDE}. For variables that change over time (such as population or cocaine crops), we use data up to 2012 when demining operations started to take place in the country \citep{local-effect}. 

\begin{figure}
    \centering
    \includegraphics[width = .5\textwidth]{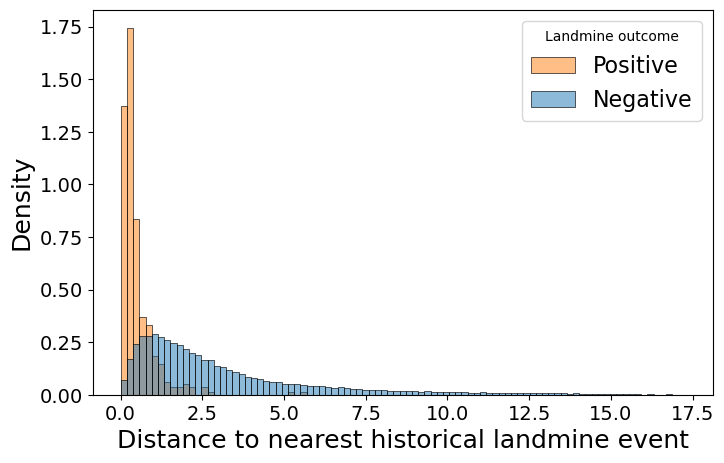}
    \caption{Distribution of distance to nearest historical landmine event for cells with positive and negative labels in our data set. This single feature is already a good predictor of landmine outcomes since positive cells strongly correlate to nearby historical landmine events and vice versa. But what about the data points when only using distance to the nearest historical landmine fails? This motivates our risk modeling methodology.}
    \label{fig:dist_old_mine}
\end{figure}

As a special case of a remnant of war indicator, we also include the historical landmine events as one of our main features \citep{departamento}, given its predictive power and paramount use in the planning of demining operations. For instance, Fig. \ref{fig:dist_old_mine} shows that cells with landmines found during demining operations (positive) are close to cells with past landmine events in a larger proportion than cells free of mines (negative). This suggests that landmines have spatial clustering patterns due to the use of several landmines to protect areas of interest instead of just a few of them. Our final dataset contains 70 tabular predictive features summarized in Tab. \ref{tab:full-importance} in Appendix.

\section{Risk Modeling}
\label{module_model}

This section outlines the methodology used to estimate the risk of antipersonnel mines. We propose an extension to the TabNet deep tabular model \citep{tabnet}, where we modify its loss function to minimize the Invariant Risk Minimization (IRM) \citep{InvariantRiskMinimization}. This modification enables the model to be robust to distribution shifts and invariant to diverse deployment environments. Additionally, we utilize SparseMax layers \citep{sparsemax10.5555/3045390.3045561} to generate global feature importance for our model.

\subsection{Preliminaries}
Here we introduce some notation and our problem formulation as a supervised classification task. Given a region, we first discretize the region into a grid of $n$ homogeneous cells defined by their location centers. For each grid cell, we collect geographical features $X_g \in \mathbb{R}^{n \times d_g}$ and historical landmine features $X_h \in \mathbb{R}^{n \times d_h}$ as our full feature set $X \in \mathbb{R}^{n \times d}$, and binary risk labels $Y \in \{0,1\}^{n}$ (0 for mine-free, 1 for mine-presence) as training data sampled from the distribution $\mathbb{P}$. Our goal is to find a function with a real value output $f_\theta: X \rightarrow [0,1]$ such that $f$ performs well on an unseen test region $(X', Y') \sim \mathbb{Q}$ that may be far away from the training region, with an arbitrary distribution shift.

\subsection{Invariant Risk Minimization for OOD Prediction}
Landmines are usually \emph{highly-concentrated} in regions where military or illegal armed group activities occurred with frequency. Demining organizations heavily rely on past landmine events to prioritize searching regions \citep{local-effect}. Due to the spatial clustered nature of this process, we encode historical landmine information with $X_h$: it is expected to find new landmines in areas with historical events nearby, and vice versa (see Fig. \ref{fig:dist_old_mine}). That is, a simple model that only uses $X_h$ should attain a good performance when evaluated on standard metrics such as the area under the ROC curve.

However, if this simple rule dominates in the learned model, the model increases false alarms in mine-free regions and fails in regions without nearby historical events. On top of $X_h$ dominated model, additional geospatial features $X_g$ can further help rule out unlikely regions and detect new landmines correctly. For naive ML models using full feature set $X$, regions without past landmine events recorded are harder to classify than places with historical landmines nearby. This can be seen as a partial correlation shift \citep{ye2022oodbench} under which the conditional distribution of landmines given covariates changes.  To alleviate the partial correlation shift, we adopt IRM \citep{InvariantRiskMinimization} where we seek to find an estimator that has a good predictive power invariant among different environments. 

In our problem, we define an ``\emph{easy}'' environment as one where landmines are found close to past events or grid cells with no historical landmines nearby have indeed negative labels. In contrast, a ``\emph{hard}'' environment is one where despite being some historical events there are no new landmines (and resources are going to be used inefficiently) or new landmines found far away from previous events (and likely missed by baseline methods leading to a latent risk to humans). We claim all landmine datasets are mixtures of ``easy'' and ``hard'' environments. Our landmine risk estimator $f_{\theta}(X)$ is penalized for applying the distance-existence rule in ``easy'' environments to ``hard'' ones, and it can generalize well on both environments with less dependence on $X_h$.

Formally, let us denote an environment by $e = (X^e, Y^e)$. For a binary classification problem, let $w$ be a dummy scalar classifier, then Eq. \ref{eq:irm} is the Lagrangian relaxation of the IRM objective which is composed by an empirical risk minimization (ERM) cross-entropy term $\ell_{\text{CE}}(f_{\theta}(X), Y) = Y\log(f_{\theta}(X)) + (1-Y)\log(1-f_{\theta}(X))$ that encourages prediction accuracy, and a regularization term that forces $f_\theta$ to be equally good across all environments $E$. The gradient term comes from Theorem 4 in \cite{InvariantRiskMinimization}, where \citeauthor{InvariantRiskMinimization} claimed that $(w^\top \Phi)^\top\nabla_w R(w^\top \Phi) = 0$ for all convex differentiable loss functions $R(\cdot)$, where $\Phi$ is the data representation $f_\theta(X^e)$.

\begin{equation}
    \textnormal{IRM}(\theta) = \min_{\theta} \sum\limits_{e \in E} \ell_{\text{CE}}(f_\theta(X^e), Y^e) + \lambda \cdot ||\nabla_{w|w=1} \ \ell_{\text{CE}}(w( f_\theta(X^e)), Y^e)||^2
    \label{eq:irm}
\end{equation}

In practice, we need to use mini-batch stochastic gradient descent to optimize Eq. \ref{eq:irm}. The distribution between easy and hard tasks is imbalanced by nature. Instead of using the same batch size for easy and hard environments, we randomly sample a mini-batch from the full train set, which is highly likely to be imbalanced, and split the batch into easy and hard environments. We then calculate the sum of gradient norms on micro-batches with different sizes respectively. If the size of mini-batch is $B$, the SGD regularization term is $\frac{\lambda}{B}(\sum||\nabla \ell (w(f_\theta(X^{\text{Hard}})), Y^{\text{Hard}})||^2 + \sum_i||\nabla \ell (w(f_\theta(X^{\text{Easy}_i})),  Y^{\text{Easy}_i})||^2)$. One $\text{Easy}_i$ and the Hard environment have the same number of samples. The sampling method automatically re-weight the importance of the two environments in the regularization term. Since we want to preserve the almost-correct rule between $X_h$ and $Y$ in easy environments, the objective will be penalized more if the invariant predictor fails on easy tasks frequently. Intuitively, the learned rule is then adjusted for hard environments using $X_g$.

\subsection{Sparse Masking for Global Feature Importance} 
\label{sec:sparse_mask_for_global_feature_importance}

\begin{figure}
    \centering
    \includegraphics[width=0.9\textwidth]{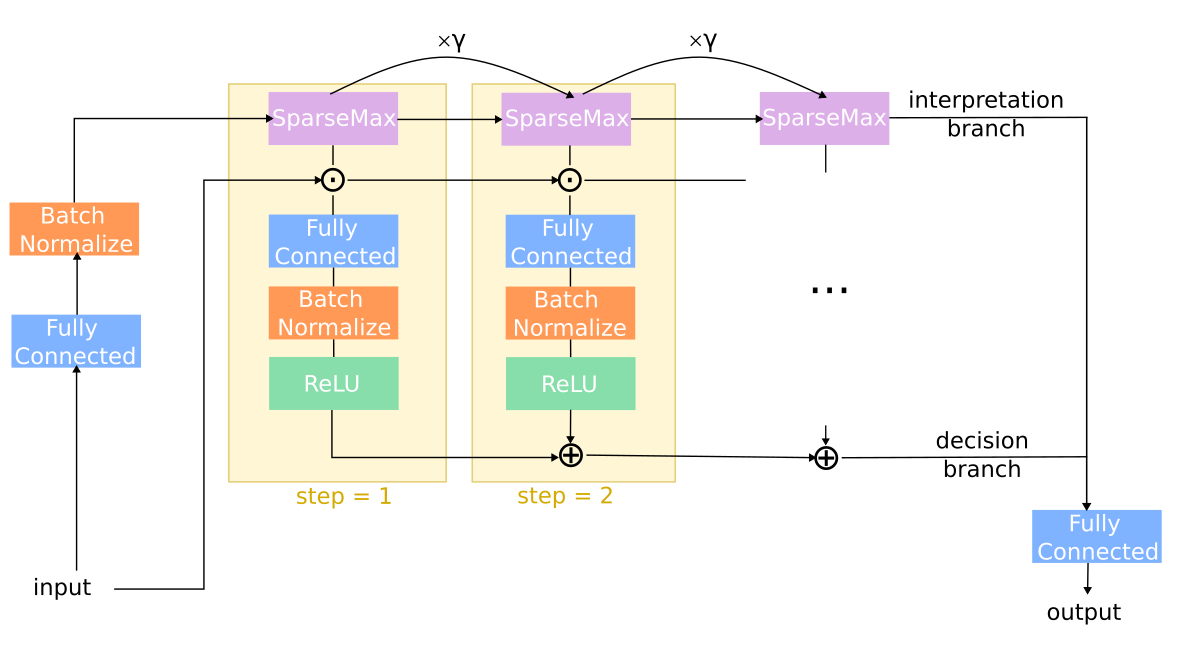}
    \caption{\textsc{\textbf{RELand}} model architecture. The model architecture is a sequential model that contains an interpretation branch that generates sparse feature masks on the top, and decision blocks at the bottom. The interpretation branch learns the sparse feature mask with one Fully Connected+Batch Normalize head and multiple related SparseMax layers. The Fully Connected+Batch Normalize+ReLU decision block can be extended with masked features. The output of each decision block is aggregated before the final Fully Connected layer.}
    \label{fig:reland_arch}
\end{figure}

An interpretable ML model will provide useful insights for decision-makers and increases trust in the system. As the first step towards the interpretation of landmine risk estimators, we start with global feature importance which can guide practitioners to identify relevant factors in landmine risk presence. We pursue a feature selection approach to construct more interpretable models. For instance, the popular LASSO model \citep{lasso1996} performs model selection simultaneously with training yielding to a sparse model. This sparse representation of the conditional distribution can be used to infer the most relevant variables to model the outcome of interest $Y$. However, the LASSO does not easily apply to modern ML problems because it fails on high-dimensional noisy datasets \citep{MEINSHAUSEN2007374relaxedlasso, zhang2010nearly, xie2009scad}. 

Recently, the deep tabular model TabNet \citep{tabnet} gained popularity due to its impressive performance on multiple tasks and its ability to capture instance-wise feature importance. Inspired by the attention transformer module in TabNet and the sparse representation in the LASSO, we design an interpretation branch as shown at the top of Fig. \ref{fig:reland_arch}. It starts from sending one data point $X_{i} \in \mathbb{R}^{d}$ into a Fully Connected (FC) layer and a Batch Normalization (BN) layer that maps input to a vector $Z_i \in \mathbb{R}^{d}$ with the same dimension. SparseMax (SM) is an activation function that normalizes the input vector to sparse probabilities \citep{sparsemax10.5555/3045390.3045561}. Formally, given a probability vector $q \in \Delta^{d-1}$, we want to find $z \in \mathbb{R}^{d}$ such that $||q - z||^2$ is minimized. There exist poly-time algorithms to find the solution to $z$, so we can efficiently project $Z_i \in \mathbb{R}^{d}$ to a sparse probability mask vector $m \in \mathbb{R}^{d}$. With learnable mask $\mathbf{M} = m \cdot \mathbf{1}^T$, we multiply $\mathbf{M}$ to $X$ element-wise to filter out less important features before passing it into the following decision branch.

\subsection{Sequential Decision Blocks}
Although Section \ref{sec:sparse_mask_for_global_feature_importance} allows us to identify the set of most important features from non-zero mask values, one-shot feature masking may lead to sub-optimal prediction for a compact model with low generalization capacity. Therefore, we leverage the sequential design in TabNet with some modifications: the decision branch receives the masked input as shown in Fig. \ref{fig:reland_arch}. Let \texttt{output} be a function of one data point $X_i \in \mathbb{R}^d$ and step $s$, and let $\odot$ be the element-wise multiplication, then the output of the full model can be defined recursively as follows:
\begin{align}
\label{eq:recursive_arch_exp}
    \textnormal{out}(X_i, 1) 
     &= \textnormal{ReLU}(\mathbf{W_{1}} [m_{1} \odot X_i])
     = \textnormal{ReLU}\left(\mathbf{W_{1}} \left[\left( \frac{1}{n}\sum_{k=1}^{n}\textnormal{SM}(\mathbf{W_{0}}[X_k])  \right) \odot X_i\right]\right) \nonumber\\
     & \vdots \\
    \textnormal{out}(X_i, s) 
    &= \textnormal{ReLU}(\mathbf{W_{s}}[m_{s}\odot X_i]) +  \textnormal{out}(X_i, s-1) \nonumber \\
    &= \textnormal{ReLU}(\mathbf{W_{s}}[\textnormal{SM}(\gamma\cdot \Pi_{j = 1}^{s - 1}m_j) \odot X_i]) +  \textnormal{out}(X_i, s-1) \nonumber
\end{align}

In Eq. \ref{eq:recursive_arch_exp}, we abbreviate parameters from FC+BN block at step $s$ with $\mathbf{W_{s}} : \mathbb{R}^{d} \rightarrow  \mathbb{R}^{latent}$, and $\mathbf{W_{0}}$ is the learnable parameters in the initial interpretation mask. $\gamma$ is a hyperparameter in $[-1,1]$ that controls how many features in previous steps are considered in the current decision block. When $\gamma \rightarrow -1$, features used in previous steps are discouraged in the current decision step and vice versa. Predictions from decision blocks are summed together and passed into an aggregation FC layer as the final prediction $\widehat{Y}_i$. 

To generate the global feature importance, we need to aggregate SM masks from multiple branches. Let $S$ be the total number of decision steps, and $\mathbf{M_{s}}_{,j}$ be the $j$-th index of $\mathbf{M_{s}}$. The normalized importance for the $j$-th feature is defined to be its contribution to the final decision output, $\sum_{s = 1}^{S} \eta_s\mathbf{M_{s}}_{,j}/\sum_{i = 1}^{d}\sum_{s = 1}^{S}\eta_s\mathbf{M_{s}}_{,i}$, where $\eta_s = \sum_{l = 1}^{latent}\textnormal{ReLU}(\mathbf{W_{s}}[\mathbf{M_{s}} \odot X])_l$. If $\mathbf{M_{s}}_{,j} = 0$ for many steps, feature $j$ leads to zero value output after ReLU activation, and thus is not important in the learned model. Feature importance is normalized to $[0,1]$ and the sum of all importance for one model is $1$.

The sequential design resembles additive modeling in Gradient Boosting Decision Trees \citep{gbdt10.1214/aos/1013203451} and ResNet \citep{resnet} skip connection mechanism \citep{10.5555/3157096.3157158}. Initial blocks capture the main correlation in the dataset, and the following blocks can use the rest of the features to learn the residuals to fit the function better \citep{NEURIPS2019_5857d68c}. Thus, it is interesting to compare selected features at different steps to verify learned importance. 

\subsection{Building Compact Model for IRM}
We want to highlight several design differences from TabNet to incorporate Invariant Risk Minimization in our pipeline. IRM suffers from overfitting and reliance on spurious features in deep overparameterized models \citep{sparseirm-pmlr-v162-zhou22e, lin2022bayesian}. Several solutions include  \cite{pmlr-v139-zhang21a} pruned networks to find the invariant subnetwork structure, and \cite{sparseirm-pmlr-v162-zhou22e} proposed sparse IRM where they learn a nondeterministic mask to filter spurious features. Instead of starting from a deep model and applying pruning methods, we directly design a lightweight and compact model to prevent overfitting on medium-scale datasets in our problem. 

First, instead of learning a feature mask for each decision step, we only learn one global feature mask before the first decision block. One obvious benefit is the removal of one hyperparameter that represents the input dimension of decision blocks. Additionally, in TabNet, re-learning extra masks on many subsets with used features can be parameter inefficient. If the input dimension of decision block is $n_a < d$, then the worst case is $\binom{d}{n_a}(n_a^2 + 3n_a)$, which can be very large for a big $d$ and just a small $n_a$. If features are used in decision blocks repetitively, our architecture encodes this information with $\gamma$ and only one FC+BN mask with $d^2+3d$ parameters. Second, in Eq. \ref{eq:recursive_arch_exp}, we mask features $X$ with mask $m_1 = n^{-1}\sum_{k=1}^{n}\textnormal{SM}(\mathbf{W_0}[X_k])$ averaged for the full train set. The mean operation enables us to learn the global feature importance across environments, instead of instance-wise feature importance (i.e., $\mathbf{M_1}_{,i} = \textnormal{SM}(\mathbf{W_0}[X_i])$) as in TabNet, where they use instance-wise masking as a de-noising technique for self-supervised representation learning. The averaged mask can provide the set of features invariantly important for both easy and hard data points, which can be easily borrowed by non-profits for demining efforts. Lastly, as the number of manually-engineered features is not large, we remove the shared decision block and extra FCs in each step to further compress the model and avoid overfitting.

\section{Interactive Interface}
\label{module_app}

\begin{figure}
    \centering
    \includegraphics[width=.9\textwidth]{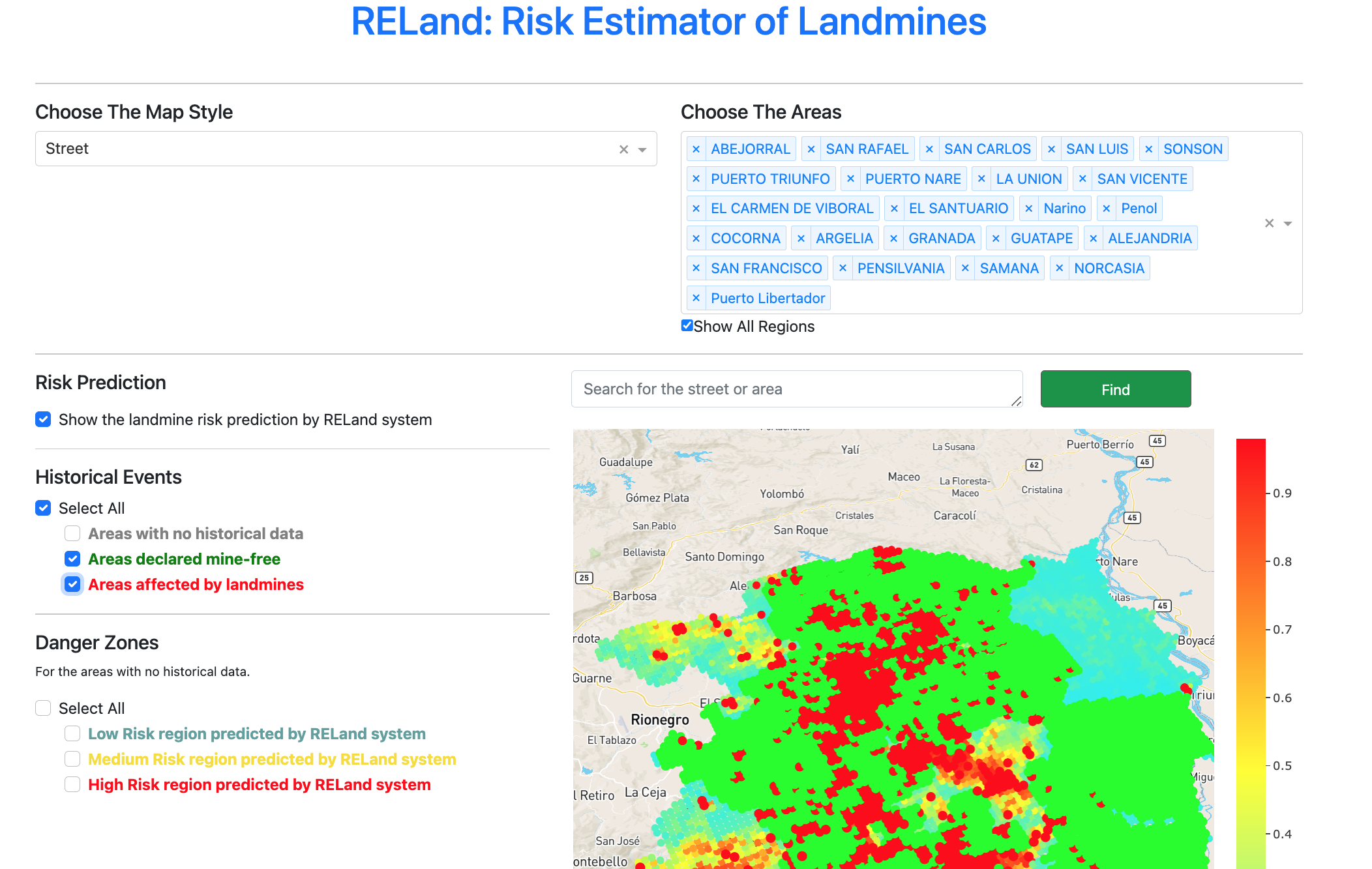}
    \caption{Screenshot of \textbf{\textsc{RELand}} interactive web interface. If the users would like to see the landmines risk prediction information of a specific region, they can search the name of the area in the search box, and the map will jump to the corresponding region. Users can freely choose the map style and areas of interest.}
    \label{fig:website}
\end{figure}

We develop a web application to help humanitarian demining organizations better understand the risk of landmine presence. Note that this web interface will not be open to the general public for ethical concerns. The map on our website shown in Fig. \ref{fig:website} contains the following three layers. 

\begin{itemize}
    \item  \textbf{Confirmed Events}: The first layer is the historical events recorded in IMSMA which is one of the main inputs for planning humanitarian demining operations.
    \item \textbf{Risk Prediction}: Users can see the prediction risk based on the predicted probability of our model fitted on all training data. Users can also select the regions to filter out prediction risks at target municipalities.
    \item \textbf{Hazardous Clusters}: We generate low, medium, and high-risk zones as clustered regions with similar levels of risk for unknown regions. We apply Local Moran's I statistic to the estimated risk scores generated by \textbf{\textsc{RELand}} to construct these risk areas as statistically significant spatial clusters at the $0.01$ level \citep{lloyd_gis}. 
\end{itemize}

Compared to Geographic Information System platforms, the lightweight interactive web interface only includes important features for demining operations and is easier to be shared within and across team members.

\section{Experiments}
\label{experiments}
In this section, we introduce extensive experiments of ML models on the constructed landmine risk dataset. In Sec. \ref{sec:validation_protocol}, we propose three validation protocols that capture the spatial correlation of out data and simulate current demining operation practices. In Sec. \ref{sec:metrics}, we describe and motivate metrics for evaluating landmine risk predictive models. Then we provide the description of a large set of baseline models for comparison study in Sec. \ref{sec:baseline_models}. We report the experimental results in Sec. \ref{sec:model_performance_comparison} showing that \textbf{\textsc{RELand}} outperforms standard and new baselines across different metrics. Finally, we discuss model interpretation in Sec. \ref{sec:interpretation}. 

\subsection{Validation Protocol}
\label{sec:validation_protocol}

\begin{figure}
    \centering
    \includegraphics[width=0.6\textwidth]{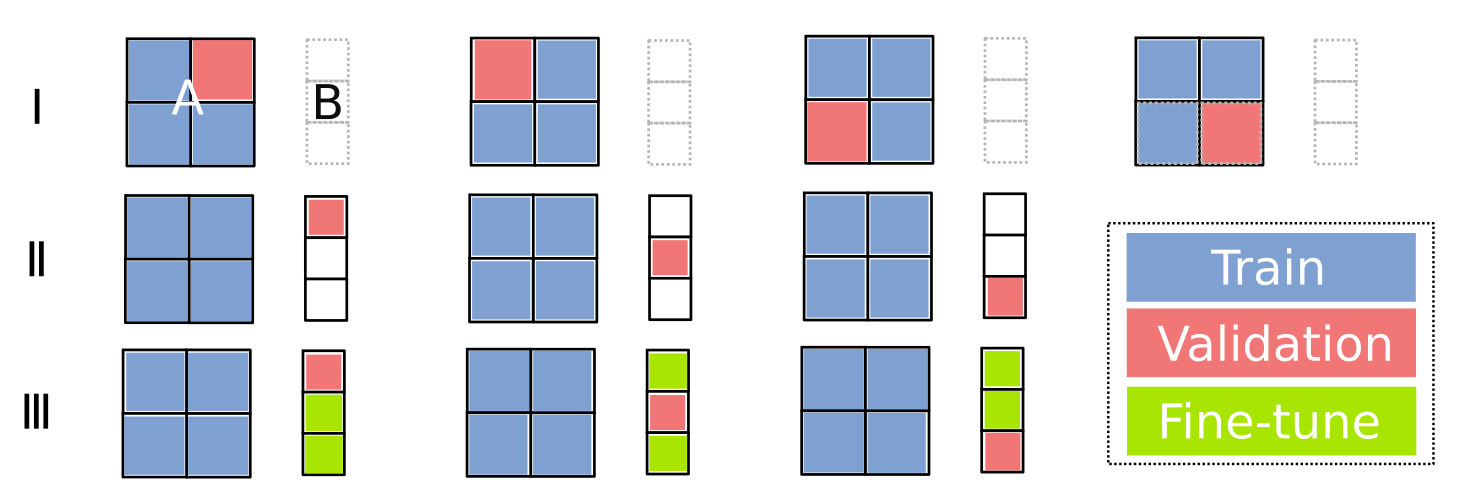}
    \caption{Visualization for three validation protocols, I = \textbf{blockCV}, II = \textbf{blockV}, III = \textbf{TransferCV}, from top to bottom. Blue blocks are used for training, red for validation, green for fine-tuning, and gray dashed blocks for datasets not used. We conduct experiments for two distant regions A and B, and report all results only on validation subregions and take the mean and standard deviation. Each A/B group provides one model for the corresponding train regions. In our experiments, A and B are two departments in Colombia, Antioquia, and Bolívar.}
    \label{fig:cv-method}
\end{figure}

Our model validation approach seeks to emulate the scenario in which the \textbf{\textsc{RELand}} system could be used, in order to correctly estimate the capabilities of the proposed methodology. Currently, the demining process is carried out at the municipal level, respecting the geographic-administrative division of the country. In this way, a humanitarian demining organization plans its operation for the whole municipality taking into account the available information and the previous experiences of other municipalities. Accordingly, we propose three evaluation methods which are visualized in Fig. \ref{fig:cv-method}.

We first use block cross-validation (\textbf{blockCV}  \citep{blockCV}) approach in which the hold-out set corresponds to all the cells in a municipality. In detail, assume we have $k$ total municipalities. We train the model with the cells of the other $k-1$ municipalities and evaluate the predictive ability of \textbf{\textsc{RELand}} in the validation municipality. This allows us to properly estimate the predictive capacity of the model when used over a new geographical region not previously demined. We repeat this process using each municipality as a validation fold, and finally report the mean and standard deviation of the relevant metrics detailed below. Using a block cross-validation strategy also allows us to incorporate the spatial dependence of the data \citep{blockCV}. Since antipersonnel landmines exhibit spatial concentration patterns, there is a correlation in the labels of contiguous cells: positive and negative cells tend to cluster together. Thus, using a standard cross-validation approach by randomly sampling a set of cells as a validation set would violate the i.i.d. assumption of the data and would result in over-confident results. We use blockCV in 15 municipalities across the Antioquia region as detailed above.

However, spatial distribution shift is one important factor that leads to poor generalization performance. Under our context, if two regions are affected by different illegal activities, new landmine risk patterns may emerge. This problem can be even more dangerous for a high-stake model targeting deployment. To test if a model is reliable at an unseen faraway region, we use a block validation approach (\textbf{blockV}) where we jointly train all data in region A, and test for each $k$ municipality in region B which is very different from A. We let A be all 15 municipalities in Antioquia and B be 5 municipalitiesin Bolívar department declared free of mines.\footnote{Cartagena de Indias, Córdoba, San Juan Nepomuceno, Santa Rosa del Sur, Zambrano} 

When data in nearby regions is available, to avoid retraining the model from scratch, it is also reasonable to fine-tune the model with data points in surrounding areas using the previously trained model in faraway regions as a warm start. We name this protocol as transfer cross-validation (\textbf{TransferCV}): given the model trained on all regions in A, we fine-tune the full model with $k-1$ municipalities in region B and test on one remaining municipality in B. The process is repeated for $k$ times.

\subsection{Evaluation Metrics}
\label{sec:metrics}
As in previous works on landmine risk prediction, we report the Area Under the ROC Curve (ROC) values to capture the overall model predictive performance. It is well known that the accuracy metric can be misleading for imbalanced data, and it only works when predicted positive and negative labels are defined at a fixed threshold. As an alternative, the ROC curve plots the true positive rate (TPR) against the false positive rate (FPR) under different thresholds of raw prediction values. If the model prediction ranks negative labels lower than positive labels, the model will get a high ROC value. However, ROC can still be misleading in an imbalanced dataset like ours, as the distribution of ground truth does not affect TPR and FPR. 

The second metric we use is the Area Under the Precision-Recall Curve (PR) more appropriate and informative for imbalanced data \citep{saito2015precision}. Compared to ROC, PR shows the trade-off between precision and recall for one model prediction. Recall is equivalent to TPR, but unlike FPR which evaluates the false positives over ground truth negatives, precision measures true positives over predicted positives, and thus is affected by the imbalance of the dataset and penalized for false alarms generated by the model.

Intuitively, in our setting, false negatives represent a higher cost in terms of human lives. Therefore, demining standards require all hazard areas to be thoroughly inspected prior to release, making the whole demining process laborious and costly. Moreover, we are not only interested in the actual risk score given to positive and negative cells but also in the ranking such scores generate, given that higher-ranked cells would potentially be prioritized in a demining operation. Better prediction for top-ranked regions may accelerate the land release, and humanitarian demining organizations can allocate the resource wisely. Thus, we also report two metrics coming from ranking learning that focus more on the top values in the resulting list. 

We use Height and Reverse Height (rHeight) proposed by \cite{rudin2009p} as metrics of how well a ranking is generated, in the sense that positive cells should be ranked higher than negative cells. Given a predicted risk score, Height refers to the average number of positive cells ranked below negative ones, and rHeight is its dual metric which is the number of negative cells above positive ones. Intuitively, if we were to clear a region sequentially following the generated risk scores ranking, rHeight tells us the average number of negative cells (false alarms) we would need to visit before fully demining the region. Therefore, this is a metric of how efficiently we could demine a geographical region of interest. An ideal classifier should minimize both of these metrics and perfectly rank positive cells above negative cells. 

Formally, 
\begin{equation}
    \textnormal{mean-Height} = \frac{1}{N}\sum\limits_{j = 1}^{N} \left[ \sum\limits_{i = 1}^{P}\mathbbm{1}(f(X_\text{p})_i \leq f(X_\text{n})_j) \right],
    \label{eq:height}
\end{equation}
\begin{equation}
    \textnormal{mean-rHeight} = \frac{1}{P}\sum\limits_{i = 1}^{P} \left[ \sum\limits_{j = 1}^{N}\mathbbm{1}(f(X_\text{p})_i \leq f(X_\text{n})_j) \right],
    \label{eq:rheight}
\end{equation}
where $P$ and $N$ are the total counts of positive and negative labels, respectively, and $f(X_\text{p}$) ($f(X_\text{n})$) is the predicted probability when the ground truth of $X_i$ ($X_j$) is positive (negative). In addition, minimizing the Height is equivalent to optimizing the initial (left) part of a ROC curve, which means we achieve a larger rate of true positives for a given rate of false positives and a more concave ROC curve \citep{rudin2009p}. 

\subsection{Baseline Models}
\label{sec:baseline_models}
We extensively compare \textbf{\textsc{RELand}} with several standard and modified machine learning models. Following previous work \citep{baseline, aspatial}, we test the performance of Logistic Regression (\textbf{LR}) and Support Vector Machine (\textbf{SVM}) on our dataset. Although \cite{aspatial} compared LR with Geospatially Weighted Regression (GWR) \citep{gwr}, we do not include it in our baseline experiments because, firstly, they showed that there are no significant gains of using GWR in this setting; secondly, vanilla GWR involves matrix inversion and multiplication for each grid, taking runtime $O(n^2d^2)$ \citep{fastGWR} and is hard to scale to a bigger dataset as ours; thirdly, GWR heavily relies on a geospatial weight matrix which cannot be transferred to distant test regions since the matrix entries will mostly become $0$ given its spatial decay component. Other models we consider include Random Forest (\textbf{RF}) \citep{random_forest}, LightGBM (\textbf{LGBM}) \citep{xgboost}, and Multi Layer Perceptron (\textbf{MLP}) \citep{mlp}, all of which are classic off-the-shelf ML models for tabular data. We also include \textbf{TabNet} \citep{tabnet} for ablation studies. We leave the hyperparameter settings and implementation details to the Appendix. 

\subsubsection{Pushed Models}
Additionally, as we use Height and rHeight as evaluation metrics, we make considerable extensions of \cite{rudin2009p} to a series of baselines and include them in our experiments for comparison. In detail, we add a relaxed version of the average Height into the loss function of the baseline ML models to build \textit{pushed} models that optimize both for predictive accuracy and for the ranking metrics simultaneously. We believe these pushed models are of interest on their own and can be used in different ranking learning settings. 

Nevertheless, the non-smooth indicator function in Height and rHeight is a challenge for optimization. We use the surrogate version proposed by \citet{rudin2009p}, called the $p$-push norm, for Height minimization: 
\begin{equation}
    \text{PNorm} = \frac{1}{N}\sum\limits_{j = 1}^{N}\left( \frac{1}{P} \sum\limits_{i = 1}^{P} \ell\left(f(X_\text{p})_i - f(X_\text{n})_j\right)^{p} \right)^{\frac{1}{p}}
    \label{eq:pnorm}
\end{equation}
where $\ell$ is a convex loss such as hinge loss or logistic loss, and $p$ is a hyperparameter that represents the strength of optimizing positive predictions. Due to its nice convexity and smoothness properties, with some minor revision, we were able to integrate $p$-norm loss into the different gradient-based optimization algorithms. In \cite{rudin2009p}, $p$-norm was originally proposed for boosted linear models, we use $p$-norm in both linear and nonlinear frameworks.

\begin{itemize}
    \item \textbf{Pushed LR and MLP}: We jointly optimize binary cross-entropy $\ell_{\text{CE}}$ and $p$-norm loss. We choose logistic loss as $\ell$ in Eq. \ref{eq:pnorm} to push predicted probabilities. The input of logistic loss is in $[-1,1]$ scale because $f$ provides probabilistic outputs. Our new loss function is therefore
    \begin{equation}
        \mathcal{L} = \ell_{\text{CE}}(X, Y) + \lambda_{p} \cdot 
        \frac{1}{N}\sum\limits_{j = 1}^{N}\left( \frac{1}{P} \sum\limits_{i = 1}^{P} \log\left(1 + \exp({f(X_\text{n})_j - f(X_\text{p})_i})\right)^{p} \right)^{\frac{1}{p}},
        \label{eq:logpnorm}
    \end{equation}
    where $\lambda_p$ controls the balance between the prediction performance of the full dataset and the positive subset. Since $p$-norm loss is convex, SGD is unnecessary for linear models like LR. We note that the $p$-norm term tends to produce extremely confident probability output.

    \item \textbf{Pushed LGBM}:
    For a GBDT classifier, we want to create a sequence of trees that minimize an objective function, typically cross-entropy. Using Taylor's expansion, the minimization problem without regularization terms is equivalent to minimizing the tree structure expression $\sum_{j = 1}^{T} \frac{G_j^2}{H_j}$, where $T$ is the data points contained in all leaves, and $G$ and $H$ are the gradient and hessian of the objective function. Using a similar loss as in Eq. \ref{eq:logpnorm}, let $\sigma$ be the sigmoid function, $\widehat{Y_\text{n}} = f(X_\text{n}), \widehat{Y_\text{p}} = f(X_\text{p})$, then we have the following gradient for the $j$-th negative and $i$-th positive data points:
    \begin{equation}
        G_{\text{n}j} = \widehat{Y}_{\text{n}j} - Y_{\text{n}j} + \widehat{Y}_{\text{n}j}(1 - \widehat{Y}_{\text{n}j})\cdot\frac{1}{P}\sum\limits_{i = 1}^{P}\sigma(\widehat{Y}_{\text{n}j} - \widehat{Y}_{\text{p}i}), \hspace{.7cm}  G_{\text{p}i} = \widehat{Y}_{\text{p}i} - Y_{\text{p}i}.
    \end{equation}

    Next, let $L = P^{-1} \sum_{i = 1}^{P} \log\left(1 + \exp({\widehat{Y}_{\text{n}j} - \widehat{Y}_{\text{p}i}})\right)^{p}$, and $H = P^{-1}\sum_{i = 1}^{P}\sigma(\widehat{Y}_{\text{n}j} - \widehat{Y}_{\text{p}i})(\sigma(\widehat{Y}_{\text{n}j} - \widehat{Y}_{\text{p}i})\widehat{Y}_{\text{n}j}^2(1 - \widehat{Y}_{\text{n}j}^2)+ 1)$, then we have an expression for the hessian of positive and negative cells:
    \begin{equation}
        H_{\text{n}j} = Y_{\text{n}j}(1 - Y_{\text{n}j}) + pL^{p-2}((p-1)G_{\text{n}j}^2+LH), \hspace{.7cm}  H_{\text{p}i} = Y_{\text{p}i}(1 - Y_{\text{p}i}).
    \end{equation}

    Note that we replace the $p$-norm term in Eq. \ref{eq:logpnorm} with $L$. Moreover, Eq. \ref{eq:logpnorm} includes the pairwise term between $X_n$ and $X_p$ and the empirical mean only reduces for part of the dataset. Worse still, we have the normalization exponent $1/p$ applied only for $N$ terms. Thus, we cannot assign the same loss value for each data point easily, so we only penalize negative data points and assume fixed positive predictions.

    \item \textbf{Pushed RF}: RF is a bag of parallel decision trees. We optimize $p$-norm greedily at each local split. Information gain subtracts weighted children entropy from parent entropy. Similarly, we set parent $p$-norm always as a constant $\log(2)$. A parent node and its left and right children create a new classifier where $p$-norm can be calculated with corresponding predicted probabilities. The split that maximizes the difference between parent $p$-norm and children $p$-norm is always chosen as we grow the trees.
    
\end{itemize}

\subsection{Model Performance Comparison}
\label{sec:model_performance_comparison}
Based on Fig. \ref{fig:cv-method}, we discuss our experiments in two parts. The first subsection involves results for validation protocol I, blockCV, where the validation region is \emph{within} the training region department. Next, we use the second subsection for validation protocol II and III, blockV and TransferCV, where we validate \emph{out-of-training} department. The latter section evaluates a harder task than the former setting.

\subsubsection{Within Department}
We compare the proposed methodology in \textbf{\textsc{RELand}} detailed in Sec. \ref{module_model} with several different models: Logistic Regression (LR), Random Forest (RF), LightGBM (LGBM), Support Vector Machine (SVM) and Multi Layer Perceptron (MLP), and pushed versions of them.\footnote{ We did not build a $p$-norm SVM model given its poor performance in the full dataset.} Motivated by the large predictive power of historical landmine events (see Fig. \ref{fig:dist_old_mine}) we also construct models using only the variable of distance to the nearest historical event (\texttt{dist\_old\_mine}), and models using all but this variable to assess how each group of features contributes to the overall performance of ML models.

\begin{table}[htbp]
\caption{Experiment result for Antioquia regions. Each entry is the mean (std) of performance on validation folds following blockCV evaluation rule. \textbf{\textsc{RELand}} is our final interpretable IRM model.}
\begin{subtable}[c]{\textwidth}
\centering
\caption{Baseline model results fitted on the single historical feature, geographical and full set of features. The best
metric for each column is bold. The best type of dataset is underlined. Note that since we average the result for different validation regions where the mine pattern varies, the standard deviation can be large. Using MLP and full datasets achieves the best score overall.}
\label{tab:baseline}

\begin{tabular}{@{}l|cc|cc@{}}
\toprule
Model                                             & ROC (↑)             & PR (↑)              & Height (↓)          & rHeight (↓)           \\ \midrule
LR-single                                         & 86.35 (11.54)       & 17.07 (10.76)       & {\ul 3.06 (3.19)}   & {\ul 226.79 (211.23)} \\
LR-geo\footnote{\cite{baseline, aspatial}} & 67.62 (18.58)       & 5.37 (8.00)         & 8.09 (6.93)         & 573.36 (440.71)       \\
LR-full                                           & {\ul 86.54 (14.06)} & {\ul 25.11 (23.54)} & 3.46 (3.09)         & 257.56 (243.23)       \\ \midrule
RF-single                                         & {\ul 85.54 (12.94)} & 15.90 (10.20)       & 3.69 (4.16)         & {\ul 331.68 (412.70)} \\
RF-geo                                            & 72.21 (11.76)       & 4.50 (4.25)         & 9.14 (9.00)         & 474.67 (406.02)       \\
RF-full                                           & 83.62 (22.49)       & {\ul 20.46 (12.14)} & {\ul 3.06 (3.41)}   & 379.58 (648.54)       \\ \midrule
LGBM-single                                       & 82.99 (16.46)       & 16.45 (9.92)        & 3.50 (3.43)         & 485.62 (730.24)       \\
LGBM-geo                                          & 64.54 (17.64)       & 11.76 (24.35)       & 11.38 (16.97)       & 812.21 (1056.11)      \\
LGBM-full                                         & {\ul 88.85 (9.19)}  & {\ul 22.03 (13.29)} & {\ul 2.82 (3.08)}   & {\ul 206.74 (206.63)} \\ \midrule
SVM-single                                        & {\ul 61.00 (33.72)} & {\ul 19.41 (19.92)} & 14.53 (21.95)       & 881.85 (1277.27)      \\
SVM-geo\footnote{\cite{baseline}}          & 48.61 (18.09)       & 1.73 (1.82)         & 15.26 (15.66)       & {\ul 821.26 (729.12)} \\
SVM-full                                          & 55.54 (16.96)       & 9.16 (10.16)        & {\ul 11.62 (12.84)} & 845.75 (767.25)       \\ \midrule
MLP-single                                        & 87.50 (10.33)       & 18.71 (11.34)       & 3.07 (3.26)         & 203.28 (184.92)       \\
MLP-geo                                           & 78.79 (12.71)       & 7.43 (9.77)         & 6.84 (6.96)         & 411.32 (361.75)       \\
MLP-full & {\ul \textbf{89.25 (11.84)}} & {\ul \textbf{27.73 (24.24)}} & {\ul \textbf{2.72 (3.25)}} & {\ul \textbf{197.88 (242.58)}} \\ \bottomrule
\end{tabular}
\vspace{2mm}
\caption{Pushed model results. The metric is bold if the model outperforms its full dataset baseline. $p = 4$ in $p$-norm loss of MLP, LGBM, RF, and $p = 2$ for LR. Using $p$-norm loss slightly improves all backbone models for different metrics.}
\label{tab:pushed_model}

\begin{tabular}{@{}l|cc|cc@{}}
\toprule
Model                                             & ROC (↑)             & PR (↑)              & Height (↓)          & rHeight (↓)              \\ \midrule
LR-pushed & \textbf{87.47 (13.91)} & \textbf{27.46 (23.31)} & \textbf{2.58 (2.57)} & \textbf{219.29 (249.89)} \\
RF-pushed & \textbf{85.13 (19.41)} & \textbf{20.51 (12.20)} & 3.09 (3.53)          & \textbf{347.99 (563.55)} \\
LGBM-pushed & 88.75 (11.69) & \textbf{27.16 (22.98)} & \textbf{2.64 (2.76)} & 263.97 (377.41) \\
MLP-pushed  & 88.38 (13.64) & \textbf{29.44 (23.21)} & \textbf{2.40 (2.61)} & 208.36 (254.98) \\
 \bottomrule
\end{tabular}
\vspace{1mm}
\caption{Top: black box models vs. Bottom: interpretable models. The best metric for each column is bold. The best model within blackbox/interpretable models is underlined. We set the step for TabNet and \textbf{\textsc{RELand}} as 2 in all tables for a fair comparison. \textbf{\textsc{RELand}} is the best interpretable model and close to the performance of the best blackbox model. Besides, it outperforms baseline and pushed models for all metrics.}
\label{tab:interpretable_model}
\begin{tabular}{@{}l|cc|cc@{}}
\toprule
Model                                             & ROC (↑)             & PR (↑)              & Height (↓)          & rHeight (↓)                  \\ \midrule
MLP-IRM & {\ul \textbf{94.15 (3.40)}} & {\ul \textbf{31.65 (24.31)}} & {\ul \textbf{2.02 (2.53)}} & {\ul \textbf{106.94 (104.94)}} \\
MLP-IRM-pushed & 92.19 (5.10)       & 30.00 (23.96)       & 2.15 (2.41)       & 142.67 (135.71)       \\ \midrule
TabNet         & 92.27 (5.54)       & 26.36 (24.91)       & 2.89 (3.02)       & 133.05 (148.32)       \\
\textsc{\textbf{RELand}} (ours)        & {\ul 92.90 (4.43)} & {\ul 29.03 (22.11)} & {\ul 2.17 (2.48)} & {\ul 132.09 (133.50)} \\ \bottomrule
\end{tabular}

\end{subtable}
\end{table}

From Tab. \ref{tab:baseline}, we observe that \texttt{dist\_old\_mine} is highly correlated with landmine risk: most landmines found after demining have close historical landmines. Based on the current demining planning practices and the histogram in Fig. \ref{fig:dist_old_mine}, it is unsurprising to see all single feature models achieving high performance in terms of ROC. Intuitively, to find most landmines it is sufficient to find an optimal threshold for \texttt{dist\_old\_mine}. 

However, these single-feature models are unable to detect any landmines found on regions without any historical data reported, where we can only use $X_g$ as the input: LR, LGBM, and MLP models using both $X_h$ and $X_g$ outperform their single-feature counterparts. Moreover, we observe that full models consistently outperform geographical feature based models used in previous literature significantly across all metrics, and there is some improvement in generalization performance from single feature models, especially in PR. Overall MLP-full model ranks the top in all baselines across all metrics.

Given the advantage of extra geographical features, we report pushed model results on the full dataset in Tab. \ref{tab:pushed_model} to see if we can further boost performance by directly minimizing Height metric that stakeholders are more interested in. In Tab. \ref{tab:pushed_model}, the gradient-based optimizer successfully minimized Height. We also see a consistent improvement in PR from baseline models. However, there is a trade-off between ROC and Height values. In \cite{rudin2009p}, minimizing $p$-norm increases the area under the left part of the ROC curve at the cost of lower ROC at the right end of the curve. Besides, although Height and rHeight represent the same intention to minimize false positives, we do not see a joint decrease of the two metrics.

Instead of using ERM as in all previously discussed baselines, combining IRM with MLP enables us to outperform all other baselines. We use \texttt{0.5km\_hist\_mines}, or the number of historical landmines within a radius of $0.5$ km to define easy and hard environments. $0.5$ km is the smallest neighborhood of historical landmines we used in our features. If \texttt{0.5km\_hist\_mines} $> 0$ but the label is 1, or \texttt{0.5km\_hist\_mines} $= 0$ but the label is 0, this data point is difficult to classify: these are either landmines found far away from historical reported events, or areas with reported events but no new landmines. Baseline and pushed models systematically miss these events. 

\begin{figure}
    \centering
    \includegraphics[width=\textwidth]{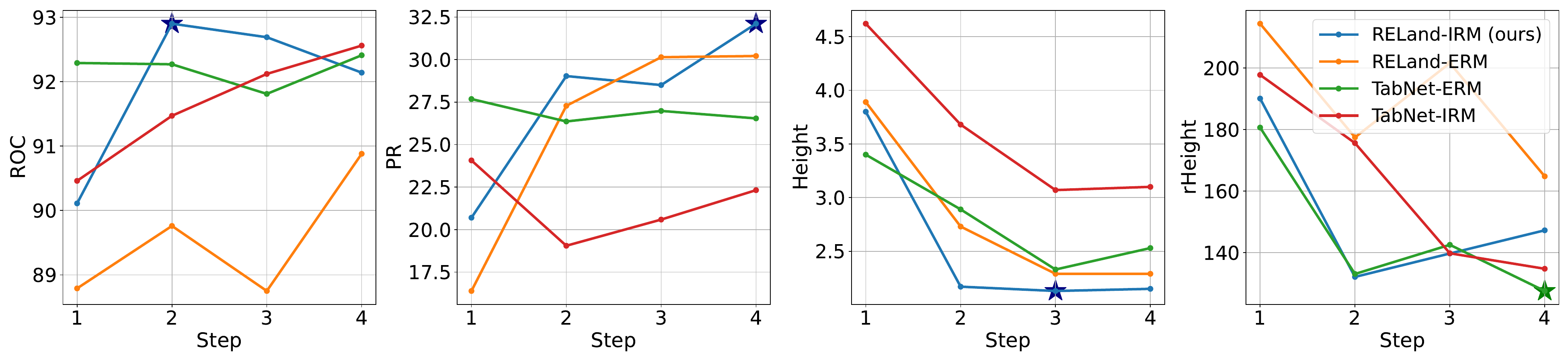}
    \caption{\textsc{\textbf{RELand}} vs. TabNet performance,  using IRM or ERM as loss function, at a different number of steps, i.e., number of decision blocks. The model that achieves the best performance is labeled for each metric. Our model (blue) achieves the best metric in most cases and ranks second for rHeight.}
    \label{fig:ablation}
\end{figure}

The experiment results for black-box models lead us in the design of our interpretable model. In Tab. \ref{tab:interpretable_model}, \textsc{\textbf{RELand}} outperforms naive TabNet across all metrics. We also observe a large reduction in the variability of our results: baseline and pushed models have three times larger ROC-standard deviation than IRM-based methods. We show the detailed performance comparison of our model and TabNet variants in Fig. \ref{fig:ablation}. The x-axis, step, is the number of decision blocks used in the model. TabNet is incompatible with IRM due to its large number of parameters. \textsc{\textbf{RELand}}-ERM is worse than our final IRM based model, which may be due to over-reliance on a wide set of distance features. \textsc{\textbf{RELand}} is mostly better than naive TabNet, except for a small discrepancy for reverse height. Interestingly, $p$-norm loss is not compatible with IRM very well. 

We conclude that \textsc{\textbf{RELand}} significantly outperforms all baseline and pushed models across all metrics in both average results and standard deviation. In other words, \textbf{\textsc{RELand}} is consistently a better model across the different municipalities (validation folds). The gains of using an interpretable component in the MLP-IRM of \textbf{\textsc{RELand}} come at the expense of a slight decrease in model performance from the black-box MLP-IRM. However, the final output of \textsc{\textbf{RELand}} and MLP-IRM is still highly correlated, with $\rho = 0.7539$. 

\subsubsection{Out-of-Department}

We test the robustness of the selected models in far away regions in Tab. \ref{tab:blockV} and \ref{tab:transferCV}. In both tables, we report our validation results in Bolívar municipalities. Based on the result using blockCV, we select several representative models including MLP, TabNet, and \textbf{\textsc{RELand}}. In general, even if our training set is in Antioquia, we observe a good performance for all models. Note that the baseline average PR here is 0.42\%, i.e., the number of positive cells over total cells, so we have already made a huge boost from the baseline. In the blockV (Tab. \ref{tab:blockV}) setting, IRM improves three metrics at the cost of PR metric, and interpretable models can further improve other metrics from black box models. More importantly, IRM reduces the variance of the metrics meaning that the models trained using IRM instead of classic ERM, have a more consistent good performance across different test regions. 

Finally, in TransferCV (Tab. \ref{tab:transferCV}) setting, interpretable models also outperform black box models. The difference between TabNet and \textbf{\textsc{RELand}} model is small, where the former has better ROC, rHeight, and the latter has better PR and Height. Finetuning improves ROC and rHeight, but not PR and Height. Overall, the proposed models seem to be robust to distribution shifts in our data and can potentially support demining operations on unseen geographical regions.

\begin{table}[htbp]
\centering

\caption{Model performance using blockV evaluation method. The best metric within the blackbox/interpretable model is bold. All models show a reasonably well performance even in this extremely imbalanced dataset. Compared to TabNet, \textbf{\textsc{RELand}} has a significant advantage on PR and Height and is close to TabNet on ROC and rHeight.}
\label{tab:blockV}
\vspace{5mm}
\begin{tabular}{@{}l|cc|cc@{}}
\toprule
Model      & ROC (↑)               & PR (↑)                 & Height (↓)           & rHeight (↓)              \\ \midrule
MLP-single & 89.98 (5.72)          & \textbf{13.13 (19.46)} & 0.62 (0.92)          & 322.20 (225.03)          \\
MLP-full   & 87.78 (9.25)          & 3.87 (6.79)            & 0.74 (0.84)          & 721.71 (1121.65)         \\
MLP-IRM & \textbf{91.79 (2.35)} & 4.74 (7.69) & \textbf{0.54 (0.78)} & \textbf{292.29 (232.03)} \\ \midrule
TabNet     & \textbf{93.90 (7.14)} & 5.89 (3.01)            & 1.00 (1.80)          & \textbf{236.85 (333.91)} \\
\textbf{\textsc{RELand}} (ours)     & 93.71 (1.15)          & \textbf{10.42 (12.42)} & \textbf{0.47 (0.69)} & 246.74 (242.89)          \\ \bottomrule
\end{tabular}
\end{table}

\begin{table}[htbp]
\centering
\caption{Model performance using TransferCV evaluation method. The best metric within the blackbox/intrepretable model is bold. Compared to blockV, fine-tuning significantly improves reverse Height at the cost of a huge drop in precision-recall. In this setting, our model is almost on par with TabNet except for reverse Height. }
\vspace{5mm}
\begin{tabular}{@{}l|cccc@{}}
\toprule
Model    & ROC (↑)       & PR (↑)               & Height (↓)  & rHeight (↓)       \\ \midrule
MLP-single & \textbf{91.13 (6.11)} & \textbf{13.25 (19.39)} & 0.62 (0.92)          & 321.05 (308.27)          \\
MLP-full & 79.01 (25.71) & 4.32 (5.09)          & 1.11 (1.25) & 1535.12 (2722.12) \\
MLP-irm    & 90.53 (6.03)          & 3.25 (4.14)            & \textbf{0.60 (0.93)} & \textbf{268.06 (161.83)} \\ \midrule
TabNet     & \textbf{94.96 (3.35)} & 2.74 (3.43)            & \textbf{0.52 (0.94)} & \textbf{117.18 (48.15)}  \\
\textbf{\textsc{RELand}} (ours)    & 94.90 (2.60)  & \textbf{2.85 (4.32)} & 0.54 (0.90) & 174.03 (158.03)   \\ \bottomrule
\end{tabular}
\label{tab:transferCV}
\end{table}

\subsection{Interpretation Analysis}
\label{sec:interpretation}
\begin{figure}
     \centering
     \begin{subfigure}[t]{0.23\textwidth}
         \centering
         \includegraphics[width=\textwidth]{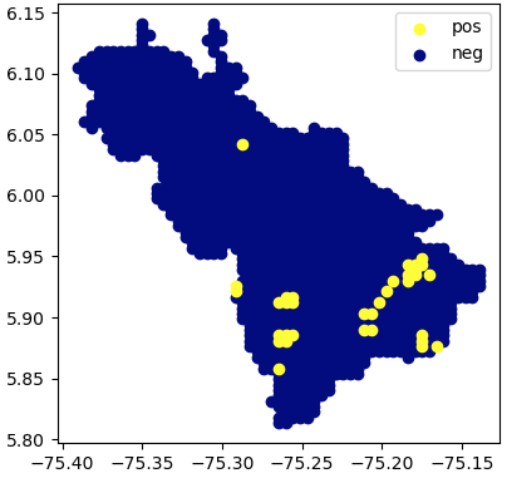}
         \caption{Ground Truth}
         \label{fig:ground_truth}
     \end{subfigure}
     \hfill
     \begin{subfigure}[t]{0.24\textwidth}
         \centering
         \includegraphics[width=\textwidth]{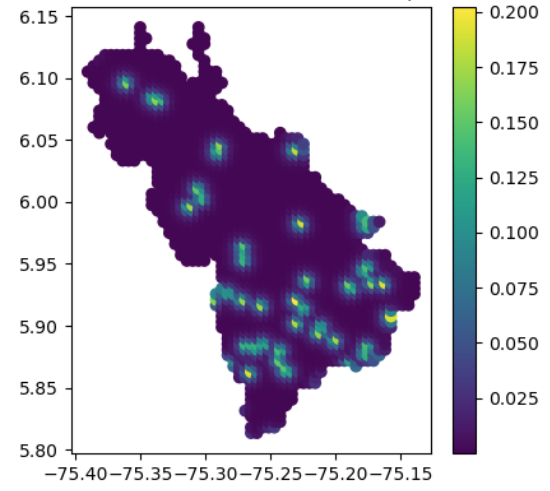}
         \caption{LR-single\\ROC: 0.911, PR: 0.161, \\Height: 3.027, rHeight: 149.8}
         \label{fig:lr_single}
     \end{subfigure}
     \hfill
     \begin{subfigure}[t]{0.23\textwidth}
         \centering
         \includegraphics[width=\textwidth]{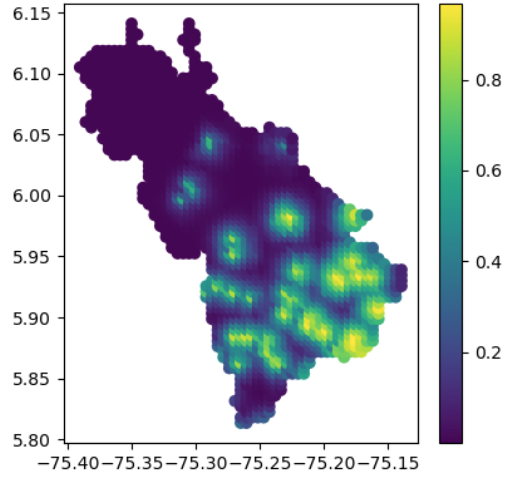}
         \caption{LR-full\\ROC: 0.927, PR: 0.275, \\Height: 3.518, rHeight: 174.1}
         \label{fig:lr_full}
     \end{subfigure}
     \begin{subfigure}[t]{0.25\textwidth}
         \centering
         \includegraphics[width=\textwidth]{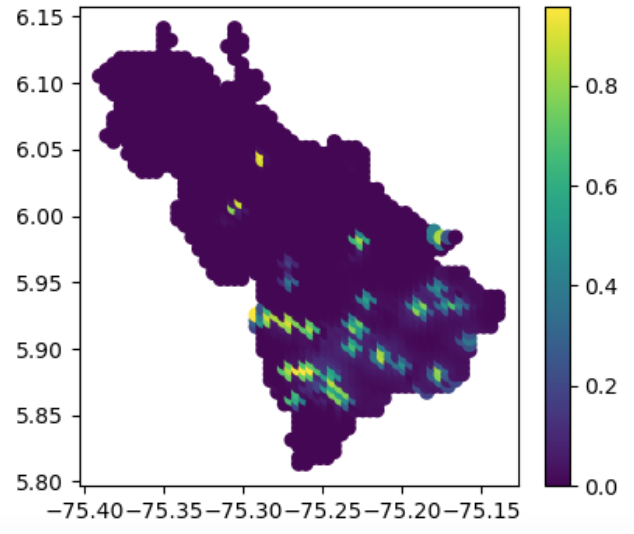}
         \caption{\textsc{\textbf{RELand}}\\ROC: 0.943, PR: 0.317, \\Height: 1.923, rHeight: 95.14}
         \label{fig:reland_full}
     \end{subfigure}
        \caption{Ground truth (a) and predicted results (b-d) from different models in El Carmen De Viboral. The color map represents the predicted risk score for grids, the x and y axis are longitude and latitude values. \textbf{\textsc{RELand}} provides the best prediction for all metrics consistently by correctly identifying precise risk clusters and removing unlikely positives.}
        \label{fig:El Carmen De Viboral}
\end{figure}

In Fig.\ref{fig:ground_truth}, we show the validation result of El Carmen De Viboral municipality as an example. Models struggle to discover landmines that are far away from the historical landmines. These events are by nature hard to predict.
Single-feature model (Fig. \ref{fig:lr_single}) simulates the current demining process well were special attention is focused on places near historical events. However, since single-feature LR is like a naive rule-based model, we can see there is a hotspot for each historical landmine. Although single-feature LR can correctly discover many landmines, it weighs each historical landmine equivalently which leads to many false alarms. In the full-LR model, this problem is alleviated when the model has the chance to reweigh grids close to historical landmines with other geographical features. However, in LR full model, when we introduce more features, the model is not confident and labels a wide range of neighboring cells as dangerous. This leads to its high Height and reverse Height metric. Due to IRM, \textbf{\textsc{RELand}} correctly makes use of extra geographical features in the full feature set to remove unlikely grids at the left top of the region and creates more localized confident predictions. As an extra bonus, compared to other models, its final prediction spreads out in the range of [0,1], making the prediction map easier for domain experts to understand as a risk score.

We then compare the feature importance of three models  under blockCV protocol: our \textbf{\textsc{RELand}} using IRM at step $= 1$, step $= 2$, and \textbf{\textsc{RELand}} using ERM at step $= 2$. Our final \textbf{\textsc{RELand}} (= IRM-2) has the least dependence on historical event features and the most dependence on geospatial features. 

\definecolor{pastelblue}{RGB}{0, 137, 252}

\begin{table}[htbp]
    \centering
    \caption{Feature Importance ([0-100] scale) comparison between \textbf{\textsc{RELand}} model variants. Features are renamed for readability. The model has different objectives (ERM/IRM) and number of steps (1/2). Feature importance values are colored based on the values. In the table, the top part represents historical event related features $X_h$, and the rest are geo-spatial features $X_g$. Our model has the smallest importance of $X_h$ and the largest importance of $X_g$. See Tab. \ref{tab:full-importance} in Appendix for the full table with all features and importance scores. }
    \label{tab:feature_importance}
\vspace{5mm}
\begin{tabular}{@{}l|cc|c@{}}
\toprule
Features                                       & IRM-1   & ERM-2   & \textbf{IRM-2 (ours)}   \\ \midrule
\quad \quad \quad \quad \quad \ 0.5km                             & \cellcolor{pastelblue!94}12.33 & \cellcolor{pastelblue!42}5.49 & \cellcolor{pastelblue!86}11.28            \\
\quad \quad \quad \quad \quad \ 2.0km                             & \cellcolor{pastelblue!88}11.53 & \cellcolor{pastelblue!53}6.91 & \cellcolor{pastelblue!6}0.81           \\
mine within 8.0km                             & \cellcolor{pastelblue!0}0.04 & \cellcolor{pastelblue!15}2.06 & \cellcolor{pastelblue!5}0.73            \\
\quad \quad \quad \quad \quad \ 16.0km                            & \cellcolor{pastelblue!0}0.00 & \cellcolor{pastelblue!10}1.34 & \cellcolor{pastelblue!5}0.73            \\
\quad \quad \quad \quad \quad \ 32.0km                            & \cellcolor{pastelblue!9}1.27 & \cellcolor{pastelblue!22}2.87 & \cellcolor{pastelblue!6}0.89            \\
distance to old mine                                & \cellcolor{pastelblue!85}11.12 & \cellcolor{pastelblue!47}6.17 & \cellcolor{pastelblue!36}4.69           \\
whether has old mine                             & \cellcolor{pastelblue!0}0.04 & \cellcolor{pastelblue!9}1.18 & \cellcolor{pastelblue!15}2.00            \\ \midrule
\textbf{Average} {$X_h$}  & 5.19 & 3.72 & \textbf{3.02} ($\downarrow$) \\
\midrule
rainfall                                      & \cellcolor{pastelblue!0}0.01 & \cellcolor{pastelblue!33}4.39 & \cellcolor{pastelblue!5}0.75            \\
land use: Ganadera                            & \cellcolor{pastelblue!0}0.02 & \cellcolor{pastelblue!5}0.74 & \cellcolor{pastelblue!29}3.87            \\
$\dots$                            & $\dots$ & $\dots$ & $\dots$             \\ \midrule
\textbf{Average} {$X_g$} & 1.01 & 1.17 & \textbf{1.25} ($\uparrow$) \\
\bottomrule
\end{tabular}
\end{table}

Since we separate ``easy'' and ``hard'' environments and reweigh the regularization term, it is easier for the model to identify landmines with selected proximity feature, so the high importance of \texttt{0.5km\_hist\_mines} is expected in \textbf{\textsc{RELand}}-IRM. This is also reflected in Fig. \ref{fig:reland_full} where we can see many small ``$+$'' that exactly represent 0.5km radius. Comparing IRM models, as we increase the number of steps, we can see a trend of decreasing feature importance in $X_h$: there is a slight decrease of feature importance of \texttt{0.5km\_hist\_mines}, and all other features in $X_h$ become close to 0. It matches our goal to make models invariant to most $X_h$. At the same time, the average feature importance of $X_g$ increases.

We suggest domain experts investigate important $X_g$ in \textbf{\textsc{RELand}} for in-depth technical analysis and Explosive Ordnance Risk Education (EORE). For example, \texttt{land\_use\_Ganadera} refers to the type of land used for raising livestock. Landmines are hard to be placed in a region that is actively used for agriculture, as livestock can trigger or detect them easily. Without IRM, \textbf{\textsc{RELand}}-ERM-2 has evenly distributed attention for all $X_h$ at step 2 and assigns high weights to some features indirectly influencing landmine events such as \texttt{rainfall}. Thus, although ERM based model already outperforms most baselines, it may provide misleading information in the learned model.

\section{Moving Towards Deployment: Pilot Field Test}
\label{field_test}

We are currently conducting a pilot field study in Colombia in partnership with a non-governmental humanitarian demining organization and an intergovernmental organization. Our partner NGO was recently allocated with two municipalities for humanitarian demining that have not been inspected before. We applied \textbf{\textsc{RELand}} to these regions to (i) build the enhanced dataset with rich geographical features, (ii) generate estimates of landmine risk by using the trained interpretable neural network IRM-based model, and (iii) present the predicted risk maps and identified hazard areas to the demining teams. We worked together with our partner NGO's on-site teams to validate the estimates generated by \textbf{\textsc{RELand}} and the priority areas identified to build an initial plan for the demining operations in the assigned municipalities.

The output for one of the municipalities included in the pilot study was validated through conversations with local communities who asked to prioritize two areas that were in turn predicted by the model as hazardous areas. Overall, the risk map generated by \textbf{\textsc{RELand}} is in line with what is expected by human experts in humanitarian demining in Colombia.  We expect to obtain initial results of the demining operations carried out in these suspect areas predicted by our model within 6 months to obtain an initial validation of \textbf{\textsc{RELand}}'s capability in real-world operations and plan the humanitarian demining operations accordingly.

Given the initial positive feedback, we believe that \textbf{\textsc{RELand}} can support crucial parts in the initial planning and the non-technical surveys in humanitarian demining operations on a large scale. Firstly, the areas predicted by the model as high risk can be prioritized by the on-site teams for additional evidence collection to confirm or cancel the suspected mined areas. Also, the model can be updated as historical landmine events in the dataset are confirmed or ruled out by the communities during non-technical surveys to update the model predictions and the predicted hazardous areas. The predictions of \textbf{\textsc{RELand}} should always be analyzed with the available information and expert knowledge to determine the actual risk of landmines in the studied area.

\section{Conclusion}
\label{conclusions}


Motivated by challenges in current demining work, this paper presents the \textbf{\textsc{RELand}} system with three major components: (i) \textbf{Dataset Enhancement:} We address the limitations of existing datasets used for landmine risk estimation by incorporating rich and relevant features that were previously overlooked in the literature; (ii) \textbf{Risk Modeling:} We propose a novel deep tabular model tailored for landmine risk estimation, and modify the loss function to minimize the Invariant Risk Minimization (IRM), ensuring that the model remains robust to distribution shifts and adaptable to different deployment scenarios; and (iii) \textbf{Interactive interface:} To validate our approach and facilitate collaboration with key stakeholders, we develop an interactive web application that allows users to explore the model's predictions. By integrating these components,  \textbf{\textsc{RELand}} offers a comprehensive pipeline for estimating the risk of landmine presence. It provides a practical and adaptable input for non-technical surveys in humanitarian demining operations, with the potential for global application and impact. 

We also admit some limitations in our work and provide possible solutions. First, our model is intended to be used by demining organizations as input in their humanitarian post-war non-technical surveys and not as an automatic real-time decision system that declares areas as safe or unsafe. Non-data-driven models might be a better choice to model the war logic given hard-to-access real-time data.  Second, our problem is not static: the risk of a mine may gradually degrade as explosive chemicals become invalid. The dataset may need to be updated for the best predictive results. The development of a model that updates its predictions according to new information collected in an incremental learning fashion, or optimizing labeling strategy is left as future research. Moreover, we foresee some ethical issues so we decided not to release our web interface to the public except for related non-profits. It would be dangerous if this information is used by people with the intention to harm communities instead of protecting them, 
undermining our goal of social good.

Landmines have caused a long-term problem in war-torn countries. We benchmarked the landmine risk prediction problem with a novel ML approach, and a newly constructed dataset, and propose an appropriate evaluation protocol. We are currently conducting a field study in Colombia in partnership with a non-governmental humanitarian
demining organization with initial positive feedback. Although acknowledging limitations, we believe \textbf{\textsc{RELand}} has the potential to support demining organizations in their demining strategies to return lands to communities, advance EORE, and help build long-lasting peace in landmine-affected countries.

\bibliographystyle{ACM-Reference-Format}
\bibliography{sample-base}

\newpage
\appendix
\section*{Appendix}
\section{Notation}
We list some commonly used notations throughout this paper. \\
\centerline{}
\begin{tabular}{p{1in}p{7.25in}}
$X_h$ & historical landmine feature \\
$X_g$ & geospatial feature \\
$X/X'$ & full train/test feature set matrix \\
$Y/Y'$ & train/test binary outcomes \\
$n$ & number of data points \\
$d$ & dimension of data, or number of features \\
$X_i$ & one single $d$-dimensional data point vector\\
$\mathbb{P}/\mathbb{Q}$ & train/test distribution \\
$f_\theta$ & supervised model parametrized with $\theta$, when the context is clear, we use the abbreviation $f$ \\
$l$ & loss function\\
$\ell_\text{CE}$ & cross-entropy loss\\
$w$ & a classifier\\
$X^e, Y^e$ & the data of an environment $e$\\
$\nabla_{w|w=1} f(w)$ & take the gradient of $f$ with respect to $w$, and set $w$ to be $1$ \\
$\Phi$ & data representation \\
$B$ & batch size \\

$\Delta^{d-1}$ & $d$-dimensional probability simplex \\
$q, z, m$ & $d$-dimensional vectors \\
$\mathbf{1}$ & all ones vector \\
$\odot$ & element-wise product \\
$\mathbf{M}, \mathbf{W}$ & $d$ by latent representation's dimension matrices \\
$\mathbf{M_s}, \mathbf{W_s}$ & the mask, parameters in the decision block at step $s$ \\
$\mathbf{M_{s}}_{,j}$ & the $j$-th index of the mask matrix $\mathbf{M_s}$\\
$\binom{a}{b}$ & the value of combination term $a$ choose $b$ \\

$k$ & total number of municipalities in our dataset \\

$P, N$ & number of positive, negative data points \\
$X_\text{p}, X_\text{n}$ & features where outcomes are positive, negative \\
$\mathbbm{1}$ & indicator function \\
$\widehat{a}$ & estimated value for $a$, when the context is clear, we may just use $a$ \\
$p$ & the exponent of $p$-norm which reflects the magnitude of pushing effect \\
$\sigma$ & sigmoid function \\
$\rho$ & Pearson correlation coefficient \\

\end{tabular}

\section{Features}
All features in our dataset are listed in Tab.\ref{tab:full-importance}. Among 70 features, 57 of them are firstly introduced in landmine risk prediction problem. We use GeoPandas to process the raw geospatial dataset stored as shapefiles.

\begin{table}[htbp]
\centering
\caption{Full raw feature importance comparison table between \textbf{\textsc{RELand}} variants. Novel features are highlighted. }
\label{tab:full-importance}
\scalebox{0.57}{
\begin{tabular}{@{}l|ccc|c@{}}  
\toprule
Features                                       & ERM-2   & IRM-2 (ours)  & IRM-1   & Data Source \\ \midrule
\textbf{0.5km\_hist\_mines}                             & \cellcolor{pastelblue!42}0.0549 & \cellcolor{pastelblue!86}0.1128 & \cellcolor{pastelblue!94}0.1233 & \cite{departamento}           \\
\textbf{2.0km\_hist\_mines}                             & \cellcolor{pastelblue!53}0.0691 & \cellcolor{pastelblue!6}0.0081 & \cellcolor{pastelblue!88}0.1153 & \cite{departamento}           \\
\textbf{8.0km\_hist\_mines}                             & \cellcolor{pastelblue!15}0.0206 & \cellcolor{pastelblue!5}0.0073 & \cellcolor{pastelblue!0}0.0004 & \cite{departamento}           \\
\textbf{16.0km\_hist\_mines}                            & \cellcolor{pastelblue!10}0.0134 & \cellcolor{pastelblue!5}0.0073 & \cellcolor{pastelblue!0}0.0000 & \cite{departamento}           \\
\textbf{32.0km\_hist\_mines}                            & \cellcolor{pastelblue!22}0.0287 & \cellcolor{pastelblue!6}0.0089 & \cellcolor{pastelblue!9}0.0127 & \cite{departamento}           \\
\textbf{dist\_old\_mine}                                & \cellcolor{pastelblue!47}0.0617 & \cellcolor{pastelblue!36}0.0469 & \cellcolor{pastelblue!85}0.1112 & \cite{departamento}           \\
\textbf{binary\_hist\_mine}                             & \cellcolor{pastelblue!9}0.0118 & \cellcolor{pastelblue!15}0.0200 & \cellcolor{pastelblue!0}0.0004 & \cite{departamento}           \\ \midrule
\textbf{Average} {$X_h$} & 0.0372 & \textbf{0.0302} & 0.0519 & -
\\ \midrule
\textbf{airports\_dist}                                 & \cellcolor{pastelblue!7}0.0092 & \cellcolor{pastelblue!14}0.0185 & \cellcolor{pastelblue!26}0.0340 & \cite{datasource9}           \\
\textbf{seaport\_dist}                                  & \cellcolor{pastelblue!5}0.0074 & \cellcolor{pastelblue!6}0.0080 & \cellcolor{pastelblue!20}0.0260 & \cite{datasource9}            \\
settlement\_dist                               & \cellcolor{pastelblue!5}0.0077 & \cellcolor{pastelblue!5}0.0077 & \cellcolor{pastelblue!20}0.0260 & \cite{datasource9}            \\
\textbf{finance\_dist}                                  & \cellcolor{pastelblue!5}0.0076 & \cellcolor{pastelblue!5}0.0073 & \cellcolor{pastelblue!0}0.0002 & \cite{datasource9}            \\
edu\_dist                                      & \cellcolor{pastelblue!24}0.0320 & \cellcolor{pastelblue!9}0.0120 & \cellcolor{pastelblue!3}0.0044 & \cite{datasource9}            \\
buildings\_dist                                & \cellcolor{pastelblue!6}0.0083 & \cellcolor{pastelblue!6}0.0082 & \cellcolor{pastelblue!0}0.0000 & \cite{datasource9}            \\
waterways\_dist                                & \cellcolor{pastelblue!5}0.0074 & \cellcolor{pastelblue!5}0.0072 & \cellcolor{pastelblue!3}0.0052 & \cite{datasource9}            \\
roads\_dist                                    & \cellcolor{pastelblue!5}0.0074 & \cellcolor{pastelblue!5}0.0074 & \cellcolor{pastelblue!27}0.0363 & \cite{datasource9}            \\
\textbf{dist\_roads\_t1}                                & \cellcolor{pastelblue!5}0.0075 & \cellcolor{pastelblue!5}0.0077 & \cellcolor{pastelblue!0}0.0003 & \cite{roads_IGAC}            \\
\textbf{dist\_roads\_t2}                                & \cellcolor{pastelblue!7}0.0091 & \cellcolor{pastelblue!28}0.0374 & \cellcolor{pastelblue!0}0.0001 & \cite{roads_IGAC}            \\
\textbf{dist\_roads\_t3}                                & \cellcolor{pastelblue!7}0.0095 & \cellcolor{pastelblue!5}0.0074 & \cellcolor{pastelblue!1}0.0014 & \cite{roads_IGAC}            \\
\textbf{coca\_dist}                                     & \cellcolor{pastelblue!5}0.0074 & \cellcolor{pastelblue!5}0.0075 & \cellcolor{pastelblue!0}0.0000 & \cite{datasource7}            \\
\textbf{dist\_pipeline}                                 & \cellcolor{pastelblue!9}0.0117 & \cellcolor{pastelblue!5}0.0073 & \cellcolor{pastelblue!33}0.0438 & \cite{datasource9}            \\
\textbf{dist\_powerline}                                & \cellcolor{pastelblue!19}0.0249 & \cellcolor{pastelblue!6}0.0078 & \cellcolor{pastelblue!5}0.0066 & \cite{datasource9}            \\
\textbf{dist\_telecom}                                  & \cellcolor{pastelblue!15}0.0206 & \cellcolor{pastelblue!25}0.0330 & \cellcolor{pastelblue!2}0.0031 & \cite{datasource9}            \\
\textbf{dist\_mining}                                   & \cellcolor{pastelblue!5}0.0077 & \cellcolor{pastelblue!5}0.0070 & \cellcolor{pastelblue!0}0.0006 & \cite{colombia-mining}           \\
\textbf{rwi}                                            & \cellcolor{pastelblue!5}0.0074 & \cellcolor{pastelblue!6}0.0084 & \cellcolor{pastelblue!11}0.0153 & \cite{datasource3}           \\
altitude                                      & \cellcolor{pastelblue!7}0.0095 & \cellcolor{pastelblue!6}0.0090 & \cellcolor{pastelblue!29}0.0385 & \cite{world-climate}           \\
\textbf{No. Víctimas por Declaración}                   & \cellcolor{pastelblue!5}0.0075 & \cellcolor{pastelblue!5}0.0077 & \cellcolor{pastelblue!48}0.0629 & \cite{datasource6}           \\
\textbf{retro\_pobl\_tot}                               & \cellcolor{pastelblue!5}0.0074 & \cellcolor{pastelblue!8}0.0107 & \cellcolor{pastelblue!0}0.0001 & \cite{panel_CEDE}           \\
\textbf{indrural}                                       & \cellcolor{pastelblue!5}0.0076 & \cellcolor{pastelblue!10}0.0140 & \cellcolor{pastelblue!0}0.0000 & \cite{panel_CEDE}          \\
\textbf{areaoficialkm2}                                 & \cellcolor{pastelblue!5}0.0074 & \cellcolor{pastelblue!11}0.0143 & \cellcolor{pastelblue!1}0.0016 & \cite{panel_CEDE}           \\
\textbf{altura}                                         & \cellcolor{pastelblue!21}0.0281 & \cellcolor{pastelblue!6}0.0084 & \cellcolor{pastelblue!4}0.0053 & \cite{panel_CEDE}           \\
\textbf{discapital}                                     & \cellcolor{pastelblue!5}0.0072 & \cellcolor{pastelblue!9}0.0127 & \cellcolor{pastelblue!0}0.0002 & \cite{panel_CEDE}           \\
\textbf{pib\_percapita\_cons}                           & \cellcolor{pastelblue!13}0.0174 & \cellcolor{pastelblue!5}0.0073 & \cellcolor{pastelblue!1}0.0014 & \cite{panel_CEDE}           \\
\textbf{hist\_mines}                                    & \cellcolor{pastelblue!14}0.0191 & \cellcolor{pastelblue!6}0.0080 & \cellcolor{pastelblue!0}0.0000 & \cite{panel_CEDE}           \\
\textbf{rainfall}                                       & \cellcolor{pastelblue!33}0.0439 & \cellcolor{pastelblue!5}0.0075 & \cellcolor{pastelblue!0}0.0001 & \cite{world-climate}           \\
\textbf{temperature}                                    & \cellcolor{pastelblue!9}0.0124 & \cellcolor{pastelblue!5}0.0070 & \cellcolor{pastelblue!18}0.0238 & \cite{world-climate}           \\
\textbf{population\_2012}                               & \cellcolor{pastelblue!5}0.0074 & \cellcolor{pastelblue!13}0.0178 & \cellcolor{pastelblue!25}0.0336 & \cite{world-pop}           \\
\textbf{soil\_texture15\_trans1}                        & \cellcolor{pastelblue!5}0.0074 & \cellcolor{pastelblue!19}0.0258 & \cellcolor{pastelblue!0}0.0000 & \cite{essd-14-4719-2022}           \\
\textbf{soil\_texture15\_trans2}                        & \cellcolor{pastelblue!12}0.0169 & \cellcolor{pastelblue!5}0.0068 & \cellcolor{pastelblue!0}0.0011 & \cite{essd-14-4719-2022}            \\
\textbf{nighttime\_lights\_2012}                        & \cellcolor{pastelblue!5}0.0074 & \cellcolor{pastelblue!9}0.0122 & \cellcolor{pastelblue!0}0.0001 & \cite{li2020harmonized}           \\
\textbf{forest\_gain}                                   & \cellcolor{pastelblue!5}0.0074 & \cellcolor{pastelblue!5}0.0070 & \cellcolor{pastelblue!0}0.0000 & \cite{doi:10.1126/science.1244693}                      \\
\textbf{forest\_loss}                                   & \cellcolor{pastelblue!5}0.0075 & \cellcolor{pastelblue!8}0.0107 & \cellcolor{pastelblue!11}0.0146 & \cite{doi:10.1126/science.1244693}           \\
\textbf{animal\_inc}                                    & \cellcolor{pastelblue!16}0.0214 & \cellcolor{pastelblue!9}0.0125 & \cellcolor{pastelblue!1}0.0015 & \cite{threatened-species}            \\
\textbf{animal\_dec}                                    & \cellcolor{pastelblue!22}0.0293 & \cellcolor{pastelblue!6}0.0079 & \cellcolor{pastelblue!37}0.0486 & \cite{threatened-species}           \\
land\_use\_Agroforestal                        & \cellcolor{pastelblue!5}0.0074 & \cellcolor{pastelblue!6}0.0088 & \cellcolor{pastelblue!0}0.0000 & \cite{datasource5}           \\
land\_use\_Agrícola                            & \cellcolor{pastelblue!10}0.0132 & \cellcolor{pastelblue!5}0.0067 & \cellcolor{pastelblue!10}0.0133 & \cite{datasource5}            \\
land\_use\_Conservación de Suelos              & \cellcolor{pastelblue!8}0.0111 & \cellcolor{pastelblue!7}0.0095 & \cellcolor{pastelblue!0}0.0000 & \cite{datasource5}            \\
land\_use\_Cuerpo de agua                      & \cellcolor{pastelblue!7}0.0094 & \cellcolor{pastelblue!19}0.0256 & \cellcolor{pastelblue!0}0.0006 & \cite{datasource5}            \\
land\_use\_Forestal                            & \cellcolor{pastelblue!13}0.0176 & \cellcolor{pastelblue!12}0.0159 & \cellcolor{pastelblue!0}0.0006 & \cite{datasource5}            \\
land\_use\_Ganadera                            & \cellcolor{pastelblue!5}0.0074 & \cellcolor{pastelblue!29}0.0387 & \cellcolor{pastelblue!0}0.0002 & \cite{datasource5}            \\
land\_use\_Zonas urbanas                       & \cellcolor{pastelblue!5}0.0077 & \cellcolor{pastelblue!13}0.0170 & \cellcolor{pastelblue!22}0.0288 & \cite{datasource5}            \\
\textbf{weather\_Cuerpo de agua}                        & \cellcolor{pastelblue!5}0.0077 & \cellcolor{pastelblue!6}0.0082 & \cellcolor{pastelblue!0}0.0001 & \cite{datasource5}            \\
\textbf{weather\_Cálido húmedo}                         & \cellcolor{pastelblue!5}0.0075 & \cellcolor{pastelblue!6}0.0086 & \cellcolor{pastelblue!0}0.0000 & \cite{datasource5}            \\
\textbf{weather\_Cálido húmedo a muy húmedo}            & \cellcolor{pastelblue!9}0.0117 & \cellcolor{pastelblue!5}0.0072 & \cellcolor{pastelblue!3}0.0045 & \cite{datasource5}            \\
\textbf{weather\_Cálido seco a húmedo}                  & \cellcolor{pastelblue!8}0.0109 & \cellcolor{pastelblue!5}0.0077 & \cellcolor{pastelblue!0}0.0004 & \cite{datasource5}            \\
\textbf{weather\_Frío húmedo a muy húmedo}              & \cellcolor{pastelblue!10}0.0136 & \cellcolor{pastelblue!5}0.0072 & \cellcolor{pastelblue!0}0.0004 & \cite{datasource5}            \\
\textbf{weather\_Frío húmedo y frío muy húmedo}         & \cellcolor{pastelblue!6}0.0081 & \cellcolor{pastelblue!37}0.0484 & \cellcolor{pastelblue!46}0.0610 & \cite{datasource5}            \\
\textbf{weather\_Frío muy húmedo}                       & \cellcolor{pastelblue!9}0.0121 & \cellcolor{pastelblue!25}0.0337 & \cellcolor{pastelblue!13}0.0175 & \cite{datasource5}            \\
\textbf{weather\_Muy frío y muy húmedo}                 & \cellcolor{pastelblue!5}0.0074 & \cellcolor{pastelblue!5}0.0073 & \cellcolor{pastelblue!3}0.0045 & \cite{datasource5}            \\
\textbf{weather\_Templado húmedo a muy húmedo}          & \cellcolor{pastelblue!5}0.0074 & \cellcolor{pastelblue!5}0.0073 & \cellcolor{pastelblue!0}0.0000 & \cite{datasource5}            \\
\textbf{weather\_Zona urbana}                           & \cellcolor{pastelblue!5}0.0075 & \cellcolor{pastelblue!5}0.0073 & \cellcolor{pastelblue!15}0.0197 & \cite{datasource5}            \\
\textbf{relief\_Cuerpo de agua}                         & \cellcolor{pastelblue!7}0.0101 & \cellcolor{pastelblue!9}0.0128 & \cellcolor{pastelblue!0}0.0000 & \cite{datasource5}            \\
\textbf{relief\_Espinazos}                              & \cellcolor{pastelblue!5}0.0074 & \cellcolor{pastelblue!5}0.0074 & \cellcolor{pastelblue!0}0.0007 & \cite{datasource5}            \\
\textbf{relief\_Filas y vigas}                          & \cellcolor{pastelblue!8}0.0107 & \cellcolor{pastelblue!20}0.0272 & \cellcolor{pastelblue!0}0.0012 & \cite{datasource5}            \\
\textbf{relief\_Glacís coluvial y coluvios de remoción} & \cellcolor{pastelblue!9}0.0119 & \cellcolor{pastelblue!11}0.0145 & \cellcolor{pastelblue!8}0.0116 & \cite{datasource5}            \\
\textbf{relief\_Glacís y coluvios de remoción}          & \cellcolor{pastelblue!10}0.0139 & \cellcolor{pastelblue!9}0.0118 & \cellcolor{pastelblue!6}0.0087 & \cite{datasource5}            \\
\textbf{relief\_Lomas y colinas}                        & \cellcolor{pastelblue!14}0.0186 & \cellcolor{pastelblue!5}0.0077 & \cellcolor{pastelblue!1}0.0017 & \cite{datasource5}            \\
\textbf{relief\_Terrazas y abanicos terrazas}           & \cellcolor{pastelblue!6}0.0084 & \cellcolor{pastelblue!5}0.0067 & \cellcolor{pastelblue!0}0.0000 & \cite{datasource5}            \\
\textbf{relief\_Vallecitos}                             & \cellcolor{pastelblue!5}0.0074 & \cellcolor{pastelblue!10}0.0137 & \cellcolor{pastelblue!18}0.0236 & \cite{datasource5}            \\
\textbf{relief\_Vallecitos coluvio-aluviales}           & \cellcolor{pastelblue!8}0.0110 & \cellcolor{pastelblue!5}0.0074 & \cellcolor{pastelblue!0}0.0008 & \cite{datasource5}            \\
\textbf{relief\_Zona urbana}                            & \cellcolor{pastelblue!5}0.0074 & \cellcolor{pastelblue!5}0.0071 & \cellcolor{pastelblue!0}0.0000 & \cite{datasource5}            \\ \midrule
\textbf{Average} {$X_g$} & 0.0117 & \textbf{0.0125} & 0.0101 & -
\\ \bottomrule
\end{tabular}
}
\end{table}
\section{Implementation Details}
LR, RF, LGBM, SVM are built mainly on scikit-learn package, and we create MLP, TabNet, RELand model on Pytorch and train with a 12GB Nvidia Titan Xp. 

\textbf{LR} has max iteration $1000$, l1 penalty with $C = 1.8791083362131904$. LR-pushed loads the LR model as a warm start, with l2 penalty and $C = 0.00012$. We use the scipy minimizer's limited memory BFGS method to solve this convex problem.

\textbf{RF} has max tree depth to be $3$. RF-pushed load RF for warm start and uses $200$ trees in the ensemble forest.

\textbf{LGBM} has learning rate $0.2955060530944092$, number of leaves $1640$, max tree depth $9$, and $1000$ boosting estimators. LGBM-pushed model has number of leaves $31$.

\textbf{SVM} uses rbf kernel and $C = 0.00001$.

\textbf{MLP} is a three layer feedforward network where the hidden layer dimension is $20$. Other hyperparameter settings are the same as those for \textbf{\textsc{RELand}}.

\textbf{TabNet} has initial learning rate $0.01$, the input dimension of attention mask is $8$ and the dimension for decision block is $8$.

\textbf{\textsc{RELand}} trains for $500$ epochs with batch size $= 2048$. We use Adam optimizer with initial learning rate $0.01$ and a decay factor $0.1$ every $75$ epochs. The initial learning rate is $0.05$ for fine tuning experiments. We take the best validation checkpoint from all epochs. The l2 regularization factor is $0.0005$. We set $\gamma$ as $-1$.

Other hyperparameters not mentioned take default values. Please refer to our codebase for more details.

\section{RELand Performance at Municipality Level}

We report the performance of \textbf{\textsc{RELand}} for all municipalities using three validation protocols. In general, our method is more reliable in regions with less imbalance.

\begin{table}[htbp]
\centering
\caption{\textbf{\textsc{RELand}} model \textbf{blockCV} results for each municipality in Antioquia. Imbalance percentage (the number of positive cells over the total number of cells) is calculated for each validation municipalitiy fold.}
\label{tab:mpio-blockCV}
\begin{tabular}{@{}l|c|cccc@{}}
\toprule
Municipality         & Imbalance\% & ROC (↑) & PR (↑) & Height (↓) & rHeight (↓) \\ \midrule
Abejorral            & 0.54        & 0.96    & 0.22   & 0.44       & 81.27       \\
Alejandría           & 0.97        & 0.84    & 0.03   & 0.45       & 92.80       \\
Chigorodó            & 0.03        & 0.83    & 0.00   & 0.41       & 493.00      \\
Cocorná              & 2.73        & 0.93    & 0.30   & 0.63       & 68.19       \\
El Carmen de Viboal  & 2.02        & 0.94    & 0.30   & 0.73       & 100.09      \\
Granada              & 11.38       & 0.89    & 0.42   & 0.62       & 78.90       \\
La Unión             & 0.15        & 1.00    & 1.00   & 1.00       & 0.00        \\
Nariño               & 3.96        & 0.93    & 0.32   & 0.83       & 83.84       \\
Sabanalarga          & 0.47        & 0.94    & 0.31   & 0.75       & 63.80       \\
San Carlos           & 1.75        & 0.92    & 0.15   & 0.43       & 228.78      \\
San Francisco        & 1.9         & 0.97    & 0.39   & 0.82       & 44.44       \\
San Luis             & 0.93        & 0.96    & 0.31   & 0.75       & 76.88       \\
San Rafael           & 2.64        & 0.94    & 0.39   & 0.69       & 79.76       \\
San Roque            & 0.12        & 0.97    & 0.02   & 0.79       & 57.50       \\
Sonsón               & 1.6         & 0.92    & 0.19   & 0.68       & 432.15      \\ \bottomrule

\end{tabular}
\end{table}

\begin{table}[htbp]
\centering
\caption{\textbf{\textsc{RELand}} performance using \textbf{blockV} validation on Bolívar department.}
\label{tab:mpio-blockV}
\begin{tabular}{@{}l|c|cccc@{}}
\toprule
Municipality        & Imbalance\% & ROC (↑) & PR (↑) & Height (↓) & rHeight (↓) \\ \midrule
Cartagena de Indias & 0.04        & 0.94    & 0.00   & 0.06       & 140.00      \\
Córdoba             & 0.08        & 0.95    & 0.27   & 0.10       & 116.50      \\
San Juan Nepomuceno & 0.04        & 0.94    & 0.00   & 0.06       & 151.00      \\
Santa Rosa del Sur  & 0.04        & 0.92    & 0.00   & 0.30       & 731.00      \\
Zambrano            & 1.94        & 0.92    & 0.24   & 1.84       & 95.21       \\ \bottomrule
\end{tabular}
\end{table}

\begin{table}[htbp]
\centering
\caption{\textbf{\textsc{RELand}} performance using \textbf{TransferCV} validation on Bolívar department.}
\label{tab:mpio-transferCV}
\begin{tabular}{@{}l|c|cccc@{}}
\toprule
Municipality        & Imbalance\% & ROC (↑) & PR (↑) & Height (↓) & rHeight (↓) \\ \midrule
Cartagena de Indias & 0.04        & 0.97    & 0.01   & 0.03       & 68.00       \\
Córdoba             & 0.08        & 0.95    & 0.01   & 0.11       & 129.00      \\
San Juan Nepomuceno & 0.04        & 0.97    & 0.01   & 0.03       & 66.00       \\
Santa Rosa del Sur  & 0.04        & 0.95    & 0.00   & 0.20       & 485.75      \\
Zambrano            & 1.94        & 0.90    & 0.12   & 2.35       & 121.42      \\ \bottomrule
\end{tabular}
\end{table}

\end{document}